\documentclass[aps,nofootinbib,superscriptaddress,pra,twocolumn,,showpacs]{revtex4-1}
\usepackage{amsfonts, amsmath, amssymb, bm, bbm, graphicx, epsfig, epstopdf}
\usepackage{braket}
\usepackage{dcolumn}
\usepackage{textcomp}
\usepackage{placeins}
\usepackage[bottom]{footmisc}
\usepackage{color}
\usepackage{soul}
\usepackage[caption=false]{subfig} 
\definecolor{forestgreen}{RGB}{34,139,34}

\usepackage[normalem]{ulem} 
\usepackage[normalem]{ulem} 
\usepackage{xcolor}
\definecolor{girlishpink}{RGB}{211, 129, 195}
\usepackage[colorlinks=true, linkcolor=teal, citecolor=blue, urlcolor=girlishpink]{hyperref}
\usepackage[capitalise,noabbrev,nameinlink]{cleveref}
\usepackage{cleveref}

\crefname{figure}{Fig.}{Figs.}
\Crefname{figure}{Fig.}{Figs.}

\crefname{equation}{Eq.}{Eqs.}
\Crefname{equation}{Eq.}{Eqs.}

\begin{document}
	\title{Unwanted couplings can induce amplification in quantum memories despite negligible apparent noise}

\author{Faezeh Kimiaee Asadi$^*$}
\affiliation{Institute for Quantum Science and Technology, and Department of Physics \& Astronomy, University of Calgary, 2500 University Drive NW, Calgary, Alberta T2N 1N4, Canada}
\author{Janish Kumar$^*$}
\affiliation{%
Department of Physics, Indian Institute of Technology Roorkee, Uttarakhand-247667, India}
\author{Jiawei Ji}
\affiliation{Institute for Quantum Science and Technology, and Department of Physics \& Astronomy, University of Calgary, 2500 University Drive NW, Calgary, Alberta T2N 1N4, Canada}
\author{Khabat Heshami}
\affiliation{National Research Council of Canada, 100 Sussex Drive, Ottawa, Ontario K1A 0R6, Canada}
\affiliation{Institute for Quantum Science and Technology, and Department of Physics \& Astronomy, University of Calgary, 2500 University Drive NW, Calgary, Alberta T2N 1N4, Canada}
\affiliation{Department of Physics, University of Ottawa, 25 Templeton Street, Ottawa, Ontario, K1N 6N5 Canada}
\author{Christoph Simon}
\affiliation{Institute for Quantum Science and Technology, and Department of Physics \& Astronomy, University of Calgary, 2500 University Drive NW, Calgary, Alberta T2N 1N4, Canada}



\begin{abstract}
Theoretical quantum memory design often involves selectively focusing on certain energy levels to mimic an ideal $\Lambda$-configuration, a common approach that may unintentionally overlook the impact of neighboring levels or undesired couplings. While this simplification may be justified in certain protocols or platforms, it can significantly distort the achievable memory performance. Through numerical semi-classical analysis, we show that the presence of unwanted energy levels and undesired couplings in an NV-center-based absorptive memory can significantly amplify the signal, resulting in memory efficiencies exceeding unity, a clear indication of unwanted noise at the quantum level. Strikingly, this effect occurs even when the apparent noise i.e., output in the absence of an input field, is negligible. We then generalize our results using semi-analytical estimations to analyze this amplification, and propose a strategy to reduce its effect. Our findings extend to memory platforms beyond NV centers; as an example, we also analyze a cavity-based rubidium memory that experiences the same issue.

\end{abstract}

\maketitle

\def\thefootnote{*}\footnotetext{These authors contributed equally to this work}\def\thefootnote{\arabic{footnote}}
 
\textit{Introduction.}--- Quantum memories provide the foundation for storing, manipulating, and processing quantum information, making them essential for advancing numerous fields, including quantum computing and communication \cite{heshami2016quantum, bussieres2013prospective, lvovsky2009optical}.  In particular, within quantum communication, quantum memories enable the efficient transmission of quantum information over long distances through quantum repeater architectures, thereby facilitating the development of quantum networks and protocols for the quantum internet \cite{kimble2008quantum, simon2017towards,jennewein2023qeyssat}. 

As of now, several quantum memory protocols such as 
electromagnetically induced transparency (EIT) \cite{phillips2001storage, fleischhauer2005electromagnetically, chaneliere2005storage,longdell2005stopped}, Raman memory \cite{michelberger2015interfacing, saunders2016cavity, guo2019high, heshami2014raman}, and
Autler-Townes Splitting (ATS) \cite{saglamyurek2018coherent, higginbottom2023memory, saglamyurek2019single}, have been proposed. Each of these protocols have their own set of advantages and disadvantages, making them preferable in different quantum platforms.  Alongside factors such as storage time, fidelity and efficiency are key figures of merit in evaluating the performance of quantum memories \cite{lei2023quantum, simon2010quantum, reim2010towards}. Efficiency refers to the probability of successfully retrieving the stored photon, while fidelity measures the overlap between the retrieved photon and the ideal target photon. Achieving a high efficiency in quantum memories is challenging due to the presence of various losses inherent in a system such as absorption and scattering. On the other hand, the presence of losses and noises, such as four-wave mixing noise and interactions with the environment, can lead to infidelities in a memory protocol \cite{michelberger2015interfacing,manz2007collisional, lauk2013fidelity}. The required efficiency and fidelity depend on the specific needs of the quantum application. However, in practice, higher efficiency and fidelity lead to better overall performance in most quantum technology implementations. To date, experimental demonstrations have achieved overall memory efficiencies exceeding 0.80 and conditional fidelities surpassing 
0.99 \cite{lei2023quantum}.




In the theoretical design of quantum memories, it is common practice to simplify the energy level structure to approximate an ideal $\Lambda$-configuration by neglecting undesired couplings to other levels \cite{gorshkov2007photon, fleischhauer2002quantum, lei2023quantum, simon2010quantum,lvovsky2009optical, heshami2016quantum}. Although this simplification may be valid for certain protocols or platforms, the reliability and effectiveness of quantum memories can be significantly impacted by unwanted couplings to both desired and undesired energy levels. As such, understanding the effect of these imperfections remains unexplored.  
Addressing and potentially mitigating these effects is therefore crucial for optimizing the performance of quantum memories.

In this paper, we present a comprehensive numerical analysis using a semi-classical approach to examine all unwanted levels and couplings within an absorptive memory based on an ensemble of NV centers. Our analysis reveals that, in the presence of all system levels and couplings, significant signal amplification leads to memory efficiencies exceeding unity. Amplification always implies noise in the quantum case \cite{clerk2010introduction, lauk2013fidelity}, which occurs in the same mode as the intended signal. However, due to the semi-classical approximation we employ, the associated noise, that is in the same mode as intended signal, is not observed, even though the amplification is captured. In fact, our semiclassical approach captures only the apparent noise, which is estimated as the output in the absence of an input field, and this noise remains negligible despite the amplification. This result shows that quantifying fidelity based solely on apparent noise is insufficient for accurately assessing memory performance. 
We then generalize our findings through a semi-analytical discussion of the effects of unwanted couplings in a 4-level system. As an example, we examine a cavity-based rubidium (Rb) memory and show that this platform can also exhibit significant amplification even in the absence of apparent noise. Our findings raise questions about whether certain previously reported memory efficiencies may have been influenced, either partially or entirely, by the amplification of the memory output, depending on how noise or fidelity was characterized experimentally.



\begin{figure}
\includegraphics[scale=1.1]{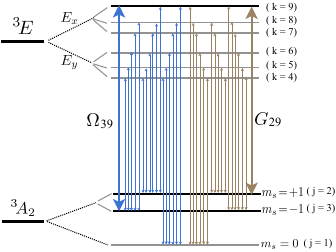}
\caption{The NV center's energy level structure, influenced by a high static electric field and a low magnetic field, is initially prepared in state $\ket{2}$. The $\ket{2}$-$\ket{9}$ transition couples to an x-polarized signal field resonant with a cavity, while the $\ket{3}$-$\ket{9}$ transition couples to a y-polarized control field. Desired couplings ($\Omega_{39}$ and $G_{29}$) are shown with thick lines, while all other couplings are undesired.} 
\label{fig:FV-levels}
\end{figure}

\textit{Numerical estimations (9-level NV center).}--- The electronic configuration of the NV center includes a ground-state triplet and six excited states.  
Here, we consider an ensemble of NV centers with the z-axis aligned along the orientation of the NV centers, and the x-axis directed along one of the reflection planes. The system is subjected to a strong static electric field and a weak magnetic field, which cause the excited states to split into the $E_x$ and $E_y$ branches \cite{heshami2014raman}. In this configuration, linearly polarized photons can couple transitions from ground states to excited levels. In our memory protocol we establish a $\Lambda$ system that consists of two ground states $\ket{+}\!=\!\ket{2}$ and  $\ket{-}\!=\!\ket{3}$ and an excited state $\ket{9}$. Here, the $\ket{2}\!-\!\ket{9}$ transition is in resonance with a x-polarized signal field that is coupled to an microcavity, assuming that the positioning of the atoms does not influence the couplings. Meanwhile, the $\ket{3}\!-\!\ket{9}$ transition resonates with a y-polarized control field. Along with the desired transitions in the system, several unwanted transitions can occur between different energy levels. \cref{fig:FV-levels} illustrate all possible transitions in the system. Here the desired Rabi frequency and cavity coupling rates are denoted by  $\Omega_{39}$ and $G_{29}$, respectively, representing the intended components. All other terms in the equations correspond to undesired couplings within the system (see \cite{SupplementaryMaterial} for couplings).
In the rotating frame the Hamiltonian can be written as \cite{SupplementaryMaterial}:
\vspace{-0.5em}
\begin{equation}
\begin{aligned}
&\hat{\tilde{H}} / \hbar =\!\sum_{k=4}^{9} \{ {\Delta_k \hat{\sigma}'_{k k}}\!- \!\hat{a} G_{1k} \hat{\sigma}_{k 1}^{\prime} e^{i\omega_{22} t}\!-\Omega_{1k} \hat{\sigma}_{k 1}^{\prime} e^{i\omega_{33} t}\\
& -\! \hat{a} G_{2k} \hat{\sigma}_{k 2}^{\prime}-\!\Omega_{2k} \hat{\sigma}_{k 2}^{\prime} e^{-i\delta t} - \!\hat{a} G_{3k} \hat{\sigma}_{k 3}^{\prime} e^{i\delta t}-\!\Omega_{3k} \hat{\sigma}_{k 3}^{\prime} \}\!-\!\!\text { H.c, }\label{H}
\end{aligned}
\end{equation}
where $\delta$ is the splitting between ground states $\ket{2}$ and $\ket{3}$, $k=4$ refers to the lowest and $k=9$ to the highest energy excited states, $G_{jk}= g_c \, g_x(j,k)$, $\Omega_{jk}=d_z g_y(j,k) E_2/2\hbar$, where $g_{x,y}(j,k)= \vec{\mu}_{jk}. \hat{x},\hat{y}/ |\mu_{jk}|$, $\vec{\mu}_{jk}=\bra{j}\vec{r}\ket{k}$, 
$\omega_2 (E_2)$ is the control frequency (amplitude), $\omega_c$ is the cavity frequency, $\epsilon$ is the permittivity of the diamond, $V$ is the cavity volume, $g_c=d_z \sqrt{\omega_c/2 V \hbar \epsilon}$ is the cavity coupling rate that satisfy the cavity cooperativity relation $C=g_c^2 N/\kappa \gamma_r$,
$\kappa$ is the cavity decay rate, $\gamma_r$ is the radiative decay rate, $d_z$ is the transition dipole moment of the zero-phonon line for the optical transition with $\lambda= 637$ nm, $N$ is the number of centers assumed to be all oriented in the same direction,
$\sigma^\prime_{kj}=\sum_{i=1}^{N}\sigma^i_{kj}$ where
$\sigma_{kj}=\ket{k}\!\bra{j}$,  $\omega_{22}=e_{22} / \hbar$, $\omega_{33}=e_{33} / \hbar$, $e_{jj}$ is the eigenenergy of the system, and $\Delta_k=\omega_{k k}-\omega_c-\omega_{22}=\omega_{k 2}-\omega_c$ is the detunings for
the $k^{th}$ excited states with respect to the ninth level. Here, we assume that the NV centers are positioned at the maximum of the cavity field, ensuring that the coupling strength between the NV centers and the cavity mode remains approximately uniform. Otherwise, one would need to account for the effects of inhomogeneous coupling, which could reduce memory efficiency and introduce decoherence effects that degrade fidelity.

Using the Hamiltonian, the Heisenberg-Langevin equations of motion for the polarization operators can be derived. Since these equations are nonlinear, we simplify the analysis by making a semi-classical approximation and treating the operators as atomic polarizations (see \cite{SupplementaryMaterial} for equations of motion). Our analysis thus neglects quantum noise and higher-order interactions. Nonetheless, as we will see below, the impact of amplification, which introduces additional noise at the quantum level, remains evident in the memory efficiency.

\begin{figure}
\includegraphics[scale=0.57]{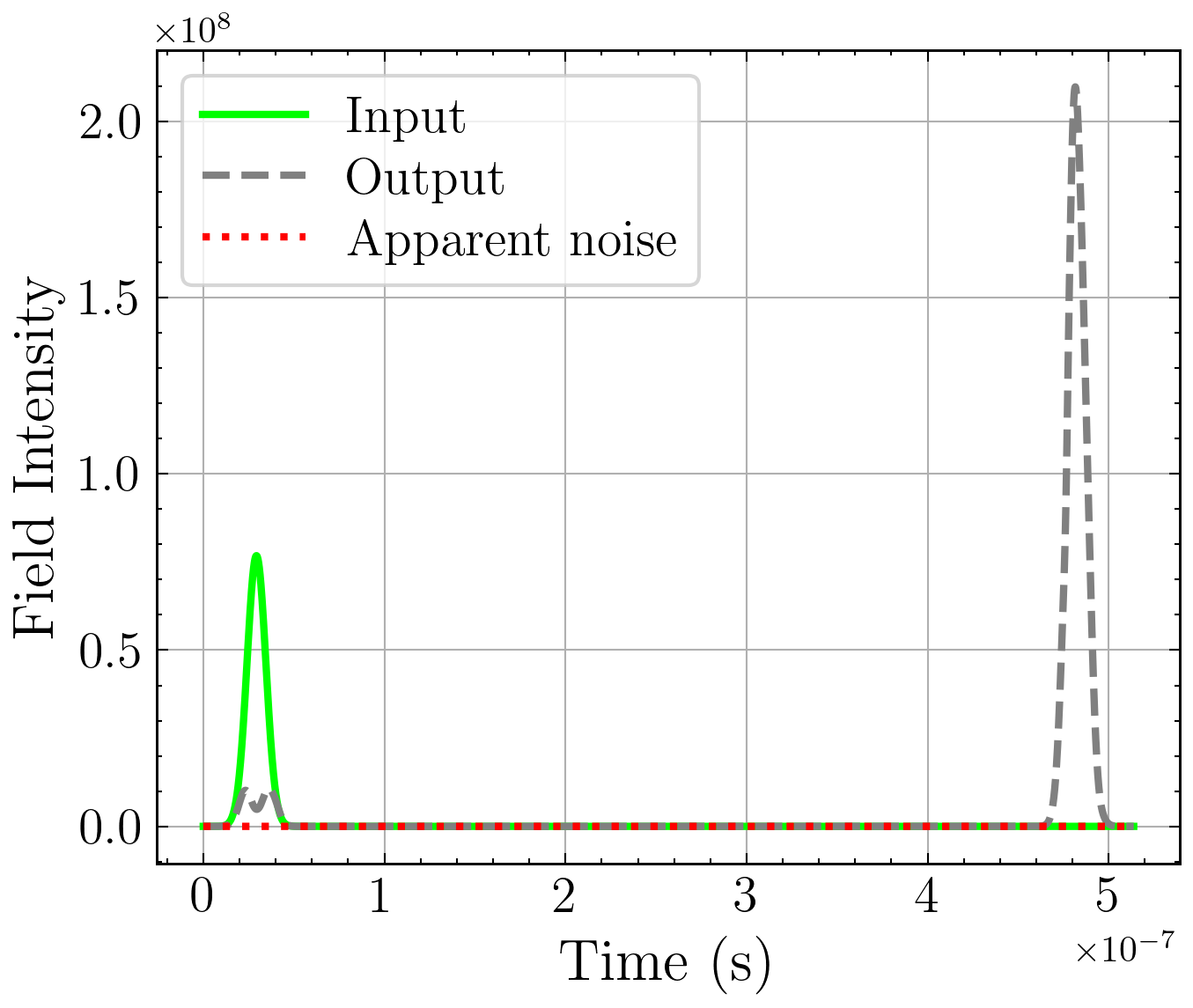}
    \caption{Storage and retrieval of the input pulse are shown as a function of time. We observe no output in the absence of an input, i.e. no apparent noise, corresponding to an apparent fidelity of unity. Here, we assumed a storage time of 455 ns to achieve maximum apparent efficiency (additional information can be found in \cite{SupplementaryMaterial}). The other parameters used are $N = 155$, $T = 2$ K, $\gamma_e = 1$ GHz, $\gamma_s = 0$, cavity $Q$-factor of 7100, signal duration $t_{\text{FWHM}} = 17.30$ ns, and a cavity volume scaling factor of 2.4. The amplitude of the first (second) control field used for storage (retrieval) is $\text{amp}_1 = 4.3$ ($\text{amp}_2 = 6$). The energy shifts in the ground state (gs) and excited state (es) due to external electric and magnetic fields are $E^{gs}_{x}=3.4$ MHz, $E^{es}_{x}=120$ GHz, $E^{gs}_{y}=E^{es}_{y}=0$,  $B^{gs}_{z}=9.9$ kHz, and $B^{gs}_{z}=10$ kHz \cite{heshami2014raman, doherty2012theory, doherty2013nitrogen}. }
    \label{amplification}
\end{figure}
To estimate the total memory efficiency, we consider a 9-level system interacting with control and signal fields considering all desired and undesired couplings that might happen. We assume initially all
NV centers are in the state $j\!=\!2$, setting $\sigma_{22}\!=\!N$. Here, we represent the input and output fields as ${a}_\text{in}(t)$ and ${a}_\text{out}(t)$, respectively. In our memory protocol, the input field is a weak field intended to operate in the single-photon regime. To ensure this, we normalize the input field such that the time integral of $|{a}_{\text{in}}(t)|^2$ equals one \cite{gorshkov2007photon}. Then, we utilize the relation
$E= \int |{a}_{\text{out}}(t)|^2 dt$ to estimate the total (apparent) efficiency of the memory. 
Note that depending on the choice of the Rabi frequency of the control fields, one can estimate the apparent efficiency of the EIT or ATS memory protocols. In  principle, if we define the $f$ factor as $f=\Omega/ \Gamma(T) $ where $\Gamma(T)/2\pi$ is the temperature-dependent homogeneous linewidth, being in the EIT regime requires $f<1$ while for ATS, we need $f>1$ \cite{rastogi2019discerning}. 
It is common to estimate the fidelity of a memory based on system noise, which can be characterized by evaluating the output field in the absence of an input field. Accordingly, we estimate the memory fidelity as $F= 1-\int |a_{\text{n}}(t)|^2 dt$ where $a_{\text{n}}(t)$ represents the output field when no input field is present. Throughout this paper, we refer to this noise as apparent noise and the resulting fidelity as apparent fidelity.
\begin{figure*}%
    \centering
{\includegraphics[width=5.5cm]{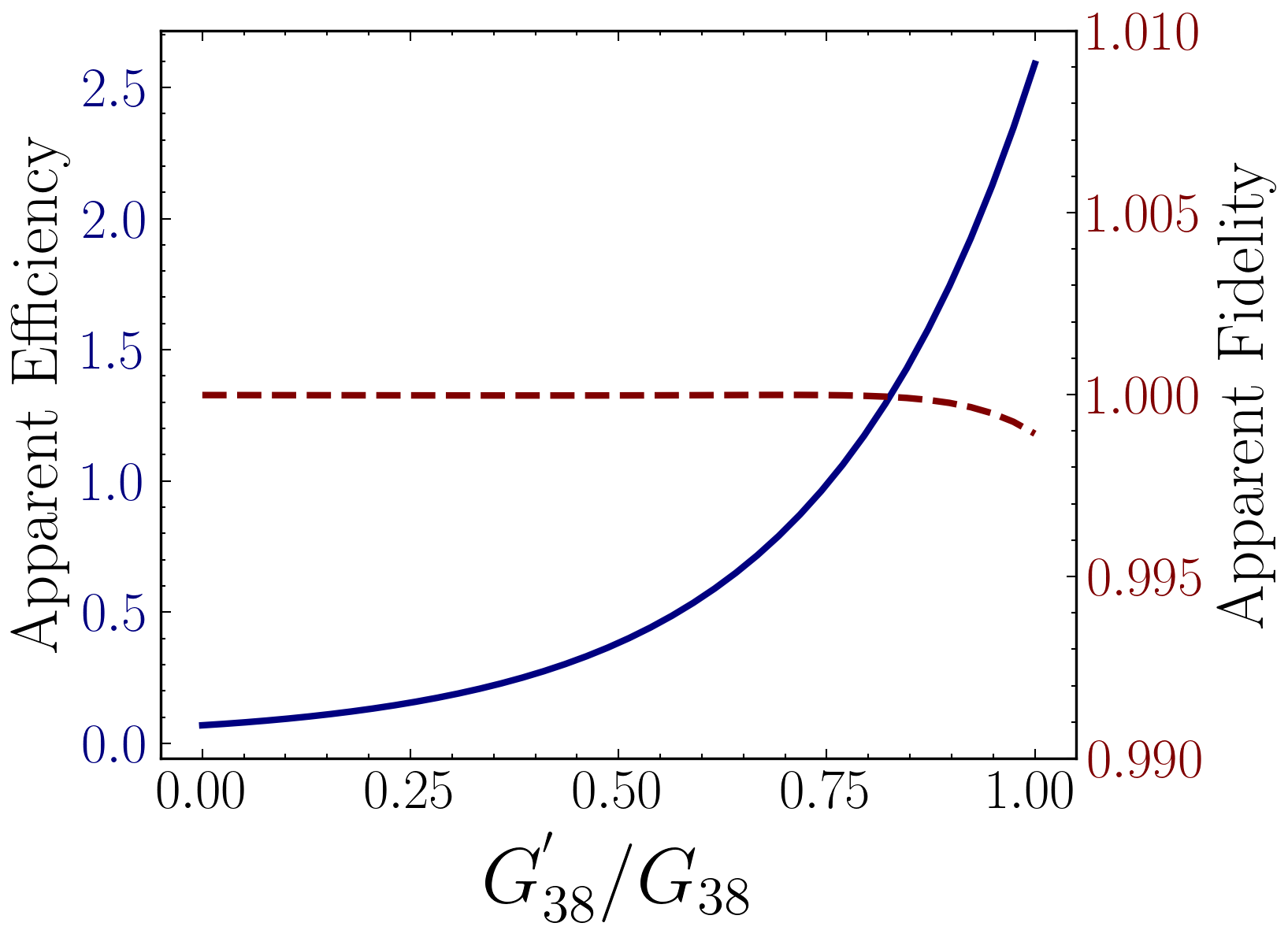} }
{\includegraphics[width=5.5cm]{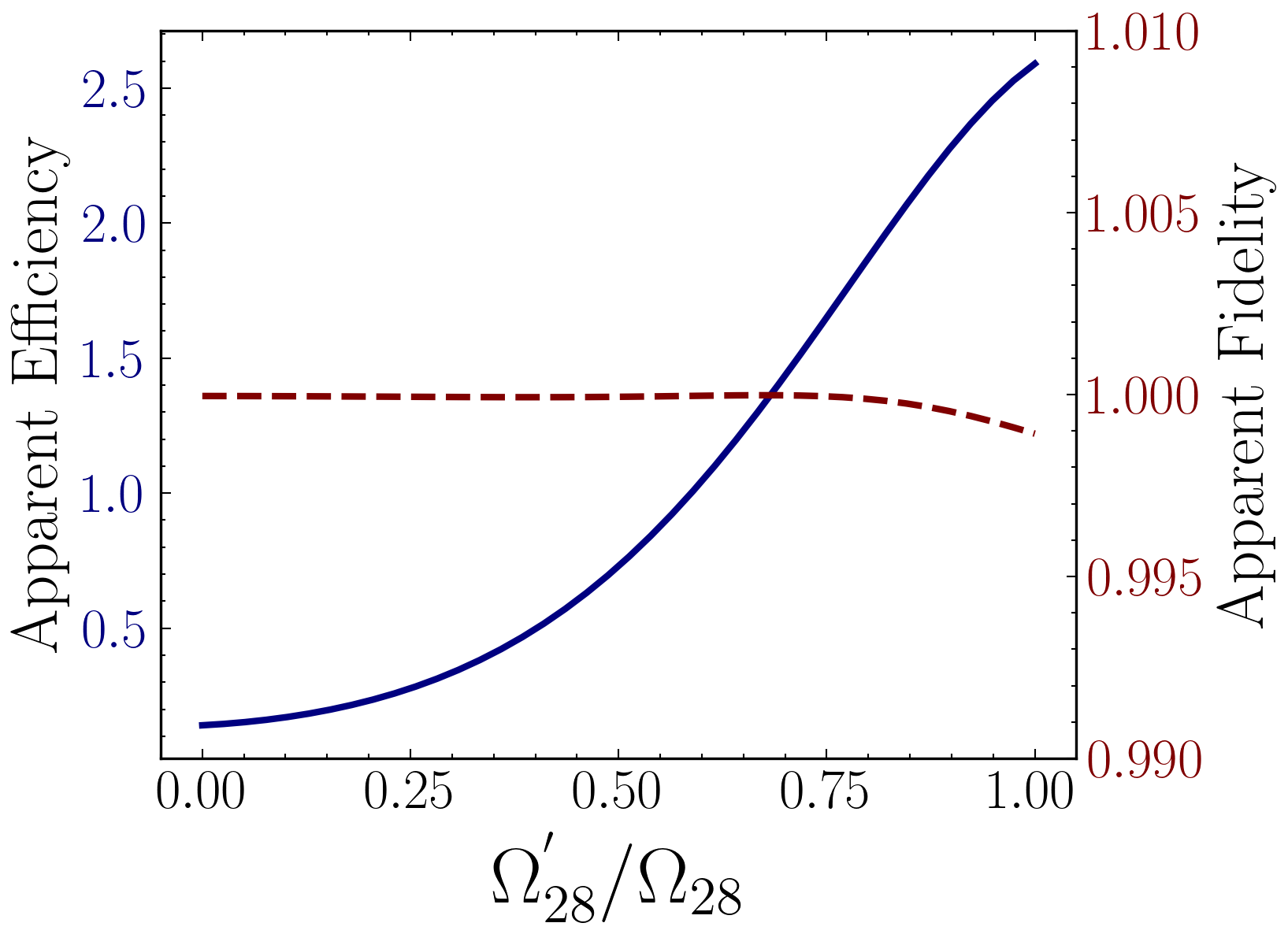} }
{\includegraphics[width=4.5 cm]{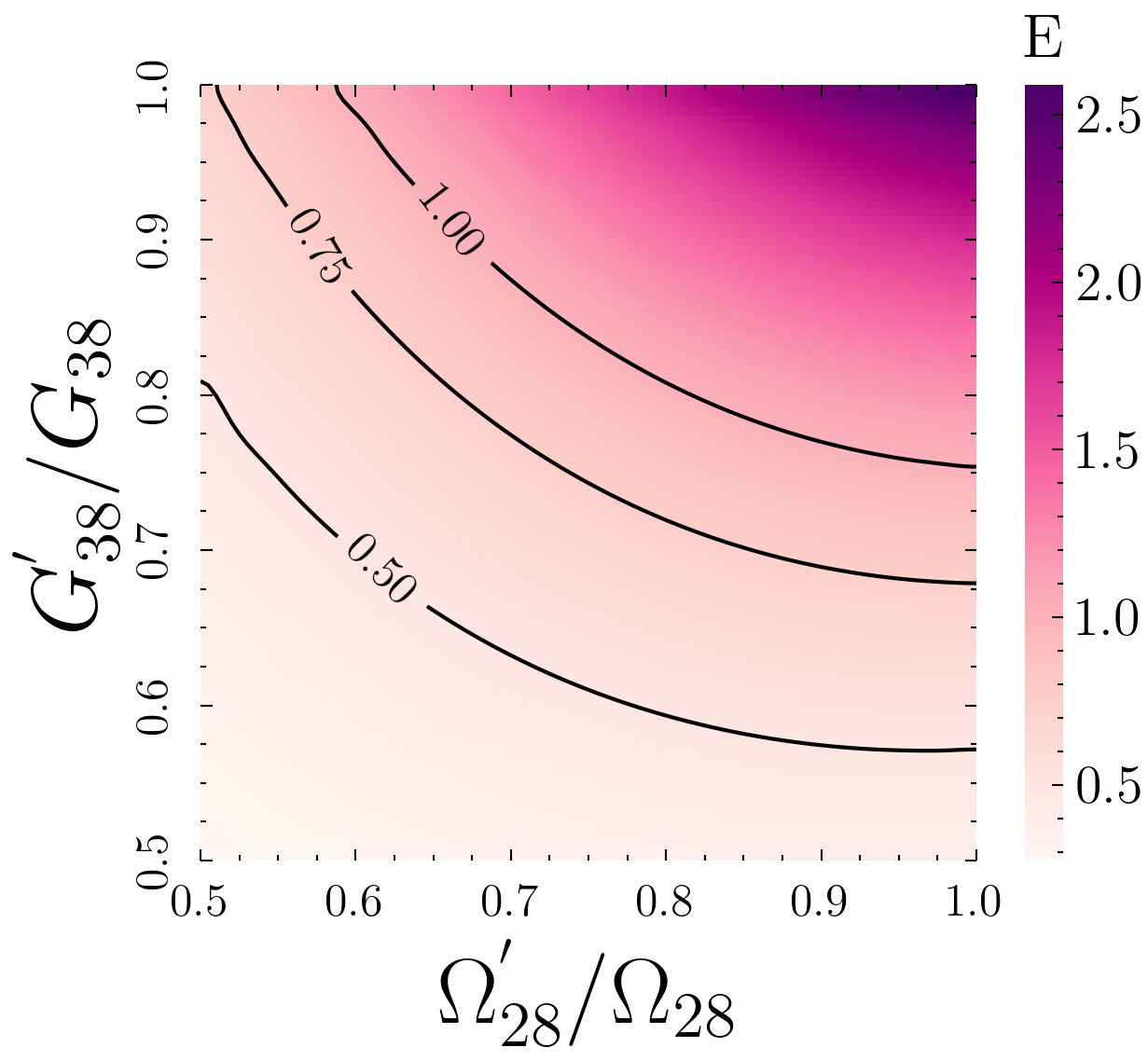} }
\caption{Numerical results: Apparent efficiency (solid line) and fidelity (dashed line) of the 9-level NV center system, including all unwanted couplings, as a function of $G_{38}$ and $\Omega_{28}$. Here, $G'_{38}$ and $\Omega'_{28}$ vary from zero to their original values of $G_{38}$ and $\Omega_{28}$, respectively, while all other couplings remain at their original values. 
The heatmap illustrates how apparent efficiency varies as a combined function of $G'_{38}$ and $\Omega'_{28}$. As shown, the presence of both unwanted couplings is essential to achieve apparent efficiencies greater than one. 
  For parameters see \cref{amplification}.}%
\label{9ln}
\end{figure*}
Considering the unwanted couplings to both the control and signal fields introduces additional linear and nonlinear terms in the equations of motion. These terms contribute to amplification, leading to an enhancement in apparent memory efficiency. Notably, as shown in \cref{amplification}, there are operating regimes where the output field intensity significantly exceeds the input field intensity, causing the efficiency to exceed unity due to amplification. This phenomenon occurs even when the apparent noise remains near zero at all times, maintaining an apparent fidelity of unity. 



Among the unwanted couplings, we identified two critical ones, $G_{38}$ and $\Omega_{28}$, whose presence is essential for achieving an apparent efficiency greater than unity in the current regime. As shown in \cref{9ln}, when these unwanted couplings are maintained at their original values, the apparent efficiency reaches 2.5. Conversely, removing either of these couplings reduces the efficiency to below unity. While the apparent efficiency is affected by amplification, the apparent fidelity remains close to unity. 

 In addition to amplification, the presence of unwanted couplings induces oscillations in the apparent efficiency as a function of storage time. This behavior arises due to a non-trivial interference between the desired and undesired couplings. It is important to note that memory output amplification can also occur in the absence of oscillations. Therefore, these oscillations are not the focus of this work (we discuss this mechanism in more detail in \cite{SupplementaryMaterial})


In order to get more insight into the amplification mechanism, we consider a 4-level system comprising ground state levels $\ket{2}$ and $\ket{3}$ and excited state levels $\ket{8}$ and $\ket{9}$, where only $G_{38}$ and $\Omega_{28}$ are present as unwanted couplings i.e., $\Omega_{29}=\Omega_{38}=G_{39}=G_{28}=0$. For simplicity, we further set $\sigma^\prime_{38}(t)=0$ and assume that almost all NV centers, initially prepared in the ground state $\ket{2}$, remain in this state at all times, i.e., 
$\sigma^\prime_{22}(t)=N$ \cite{gorshkov2007photon}.
Despite the simplifications, the system still experiences significant amplification while the apparent fidelity remains exactly unity (see Supplementary \cite{SupplementaryMaterial}). This result shows that the slight deviation of apparent fidelity from unity in the 9-level system, as shown in \cref{9ln}, can be attributed to the presence of additional unwanted couplings illustrated in \cref{fig:FV-levels}. In the following, we analytically discuss this simplified 4-level system.




\begin{figure}
\includegraphics[scale=0.42]{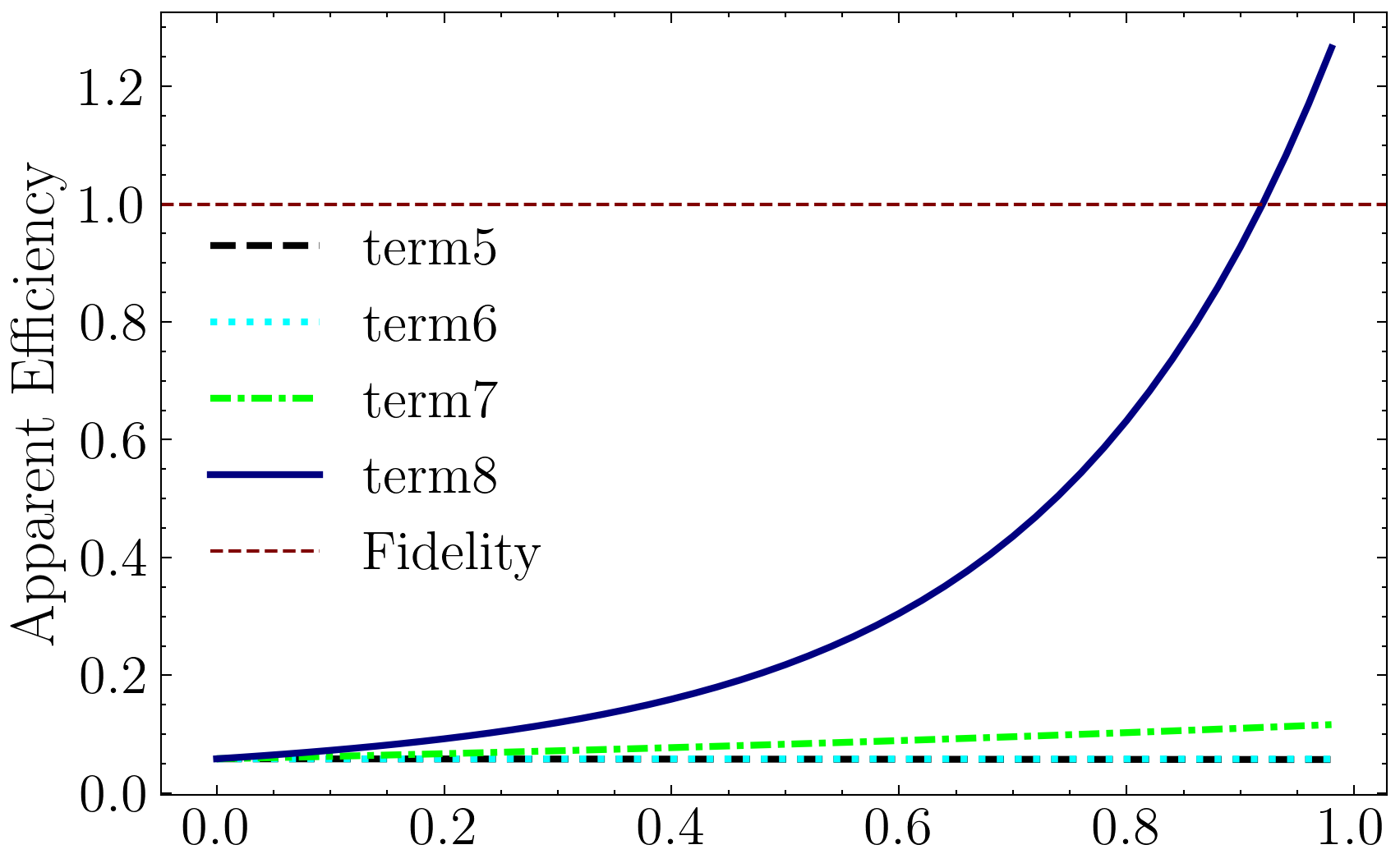}
    \caption{Apparent efficiency of the 4-level system as a function of different terms in \cref{sigma}. Here, terms 1 and 3 are retained, while terms 4 to 8 are individually varied from zero to their original values along the x-axis (for simplicity, we set $\gamma_s=0$). Additionally, in all cases presented here, the apparent fidelity remains constant at unity.
    All parameters are consistent with those used in \cref{amplification}, except for $\text{amp}_1 = 4.3$ and $\text{amp}_2 = 1.5$, which are adjusted to ensure that the assumption of adiabatic elimination remains valid. }
    \label{terms}
\end{figure}
\textit{Semi-analytical estimations (4-level system).}--- For the simplified 4-level system discussed above, the equation of motion for $\sigma^\prime_{32}$ can be written as follows
\begin{equation}
\begin{aligned}
&\dot{\sigma}^\prime_{32}(t) + \gamma_s \sigma^\prime_{32}(t) + \!
\frac{\sqrt{2\kappa} \,N\,  G^*_{29}\, \Omega_{39}  \,a_{\text {in}}(t)}{\alpha} + \!
\frac{\kappa\, |\Omega_{39}|^2 \,\sigma^\prime_{32}(t) }{\alpha}\\&+ \frac{|G_{29}|^2 \beta(t) \beta^\dagger(t)  \sigma^\prime_{32}(t)}{(\gamma_d + \gamma_e)\alpha^2} + 
\frac{|G_{38}|^2 \beta(t) \beta^\dagger(t) \sigma^\prime_{32}(t)}{(\gamma_d + \gamma_e - i \Delta_8) \alpha^2}\\&+ 
\frac{e^{2 i t \delta} \sqrt{2 \kappa}\,N(\gamma_d + \gamma_e)  \, G_{38} \,\Omega^*_{28} \,a_{\text {in}}(t) }{(\gamma_d + \gamma_e - i \Delta_8) \alpha}\\&-
\frac{e^{2 i t \delta} \,N  \, G_{38} \,\Omega^*_{28}  \Omega_{39}\, G^*_{29} \,\sigma^\prime_{23}(t)}{(\gamma_d + \gamma_e - i \Delta_8) \alpha}
=0, \label{sigma}
\end{aligned}
\end{equation}
where $\alpha=(\gamma_d + \gamma_e)\kappa + |G_{29}|^2 N$, $\beta(t)=\sqrt{2 \kappa}(\gamma_d + \gamma_e) \,a_{\text {in}}(t) - \Omega_{39}\, G^*_{29} \,\sigma^\prime_{23}(t)$,
$\gamma_s$ is the spin inhomogeneous broadening, $\gamma_{e}$ is the optical inhomogeneous broadening, and $\gamma_d$ is the decoherence rate of the optical transitions. In deriving the above equation, we employed the adiabatic elimination of the cavity mode, $\sigma^\prime_{39}$, $\sigma^\prime_{29}$, and $\sigma^\prime_{28}$ \cite{gorshkov2007photon} (see \cite{SupplementaryMaterial}). Consequently, the following results apply to memory protocols based on the adiabatic elimination of absorption, such as EIT \cite{rastogi2019discerning}. 
Solving \cref{sigma} analytically is not feasible; however, a numerical solution can be employed to quantify the system's performance (see \cite{SupplementaryMaterial}). 
In \cref{sigma}, terms 1 and 3–5 represent the desired contributions, while terms 2 and 6–8 correspond to the undesired components. \cref{terms} illustrates the impact of different terms in \cref{sigma} on the apparent efficiency and fidelity of the memory. As shown, the amplification in the memory output is primarily driven by term 8, as the presence of this term alone is sufficient to produce an apparent efficiency greater than unity.
Thus, if we further simplify \cref{sigma} by retaining only terms 1 and 8, we obtain the amplification equation, which provides a qualitative understanding of the amplification process. This equation is given by:
\vspace{-0.2em}\begin{equation}
\dot{\sigma}^\prime_{32}(t) -
\frac{e^{2 i t \delta}\, \,N\, G^*_{29}\,\Omega_{39} \,G_{38} \, \,\Omega^*_{28}  }{(\gamma_d + \gamma_e - i \Delta_8)\alpha}\,\sigma^\prime_{23}(t)
=0. \label{sigmashort}
\end{equation}
The solution to this equation involves an exponentially growing term of the form 
$
\text{exp}\,(i\delta+\sqrt{b^2/(\Gamma^2+\Delta_8^2)-\delta^2})t
$ where $\Gamma=\gamma_e+\gamma_d$, and 
\vspace{-0.2em}
\begin{equation}
b=\frac{NG^*_{29}\,\Omega_{39} \,G_{38} \, \,\Omega^*_{28}}{\alpha}.\end{equation}
The form of $b$ shows that this amplification originates from a complex four-wave mixing (FWM) process caused by unwanted coupling between the ground and excited states. 
It is important to note that this FWM process involves simultaneous contributions from unwanted couplings to both the control  (i.e., $\Omega_{28}$) and signal (i.e., $G_{38}$) fields, which 'reinforce' each other and result in significant amplification. Therefore, considering only the unwanted coupling to the control field, a common practice in many theoretical studies aiming to assess the impact of FWM noise \cite{lauk2013fidelity}, can overlook the potential for substantial amplification effects on memory performance. Given that the noise discussed here is a new form of FWM, methods previously developed to mitigate other forms of FWM noise \cite{nunn2017theory, thomas2019raman} and to detect its presence \cite{ thomas2019raman} may still be applicable in the present context. However, accurately evaluating the effectiveness of these methods requires a full quantum treatment.
Enhancing memory efficiency generally increases both true efficiency and amplification; however, the latter grows exponentially.  
To mitigate amplification, it is beneficial to increase
$\delta$ (ground states splitting) and $\Delta_8$, besides minimizing the unwanted couplings.

  \begin{figure}
    \centering
{\includegraphics[width=6.6cm]{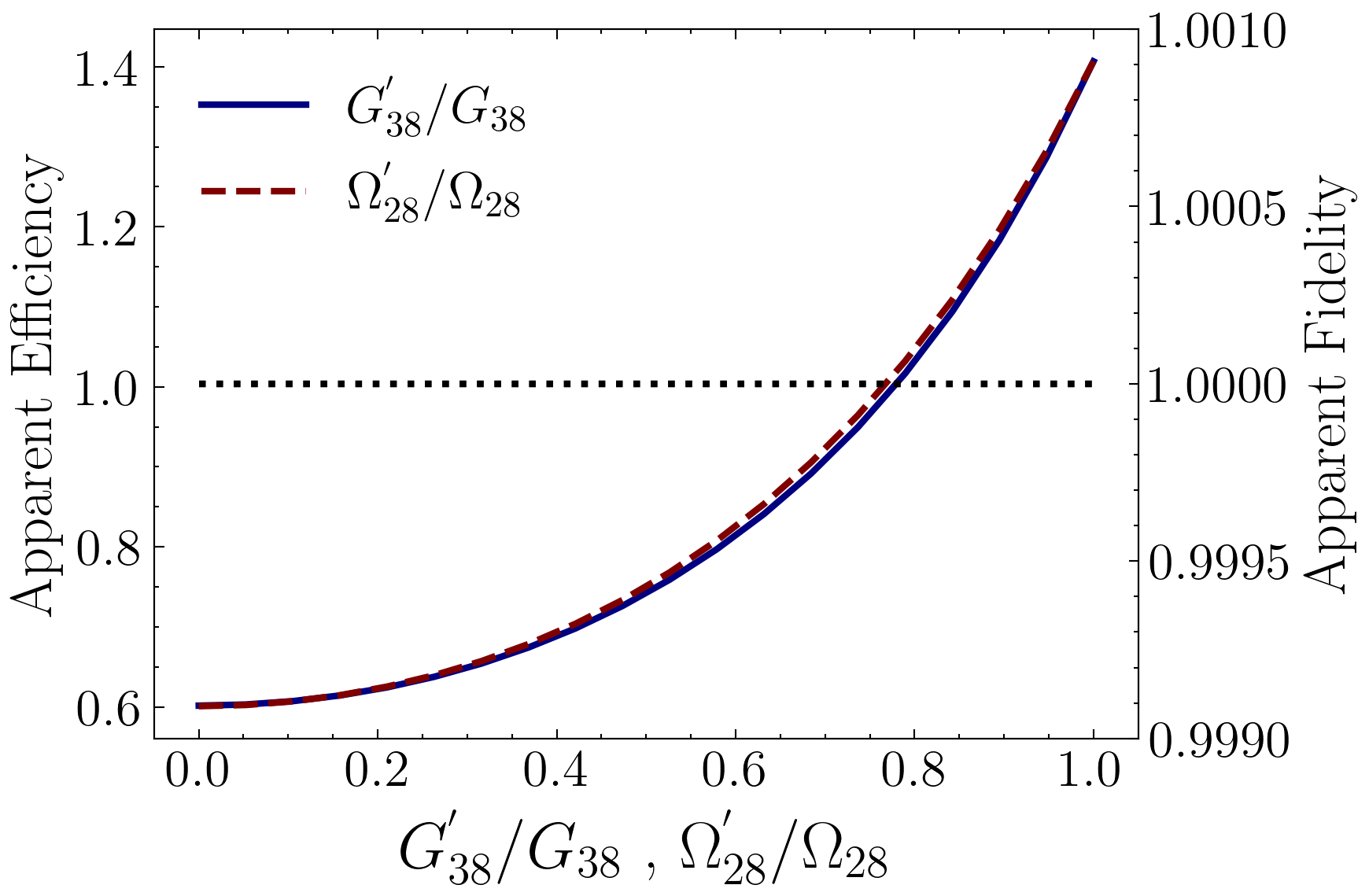} }
    \caption{Numerical results: Apparent efficiency (solid line) and fidelity (dotted line) of the cavity-based $^{87}$Rb memory as a function of $G_{38}$ and $\Omega_{28}$, with all other unwanted couplings  set to zero. 
   Here, we set $N = 250$, $Q = 7100$, $t_{\text{FWHM}} = 17.30$ ns, a cavity volume scaling factor of 1.5, a storage time of 91 ns, and $\text{amp}_1 = 0.05$ and $\text{amp}_2 = 0.1$ (see \cite{SupplementaryMaterial,steck2001rubidium} for more information).}%
    \label{Rb}
\end{figure}

It should be noted that the amplification-related noise discussed in this paper is not limited to the NV center. Our findings are also relevant to solid-state-based memories with weak selection rules and atomic memories. In particular, previously reported impressive memory efficiencies in atomic memories may have been significantly influenced by amplification. For instance, consider a hypothetical cavity-based $^{87}$Rb memory with a $\Lambda$ configuration consisting of $\ket{2}\!\equiv \!\ket{g}\!=\!\ket{5S_{1/2}, F\!=\!1}$ and $\ket{3}\!\equiv \!\ket{s}\!=\!\ket{5S_{1/2}, F\!=\!2}$ as the ground states, and $\ket{9}\!\equiv \!\ket{e}\!=\! \ket{5P_{1/2}, F\!=\!2}$ as the excited state (see \cite{SupplementaryMaterial} for more details). As shown in \cref{Rb}, the presence of the unwanted couplings $\Omega_{28}$ and $G_{38}$ through the unwanted level $\ket{8}\!\equiv\!\ket{e'}\!=\!\ket{5P_{1/2}, F\!=\!1}$ leads to an apparent efficiency greater than one, accompanied by an apparent fidelity of unity in the absence of other unwanted couplings.
Among atomic systems, hot vapor-based memories \cite{ma2022high, guo2019high, wu2025ai} are particularly susceptible to the amplification effect discussed here. To mitigate this effect, employing appropriate polarization selection rules may be beneficial. However, their effectiveness often depends on optically pumping the system into specific Zeeman levels \cite{PhysRevA.90.033823}, a technique that has not been implemented in most hot vapor experiments to date.

\textit{Conclusion.}--- Considering both desired and undesired couplings is essential for providing a realistic theoretical assessment of memory performance in certain platforms. In this paper, we used an NV-based memory with a 9-level system as an example to numerically demonstrate that unwanted couplings can lead to signal amplification, which inherently introduces noise in the quantum case. Our analysis revealed that, given the NV center's energy level structure, either amplification is significant or both efficiency and fidelity are too low for practical use.

To generalize our findings, we conducted a semi-analytical analysis of a 4-level system exhibiting significant amplification, examined a cavity-based rubidium memory as an example, and discussed a method to mitigate the amplification effect. Although a semi-classical approach has enabled us to capture the effect of amplification on memory efficiency, accurately quantifying the associated noise requires a full quantum treatment of the Heisenberg-Langevin equations.  This involves incorporating bosonic modes \cite{lauk2013fidelity} and accounting for all unwanted couplings through 'both' control and signal fields, not just the control field. We expect that such a treatment would further refine our conclusions.

Our results suggest that, from an experimental per-
spective, measuring noise in the absence of an input field
is not a reliable method for estimating memory fidelity
and can lead to misleading conclusions. Instead, fidelity should be evaluated using alternative approaches, such as storing and retrieving an entangled photon with high efficiency. Operating in the high-efficiency regime is crucial, as amplification-related effects often increase with efficiency due to their correlation. Another method to estimate the contribution of amplification to memory performance is by measuring the second-order intensity autocorrelation function $g^2_{out}(0)$, to probe the quantum statistics of the retrieved photon \cite{ thomas2019raman} (see also Eq. (3.23) of \cite{hird2021engineering}). However, it should be noted that $g^2_{out}(0)$ also reflects the relative contributions of other noise sources, such as imperfect optical pumping. Therefore, in order to isolate the impact of four-wave mixing, other noise sources must be characterized and quantified separately \cite{ thomas2019raman}.

This work is funded by the NSERC Alliance quantum consortia grants ARAQNE and QUINT and the NRC High-throughput Secure Networks (HTSN) challenge program.





	\bibliographystyle{apsrev4-1}
	\bibliography{ref}

\begin{thebibliography}{36}%
\makeatletter
\providecommand \@ifxundefined [1]{%
 \@ifx{#1\undefined}
}%
\providecommand \@ifnum [1]{%
 \ifnum #1\expandafter \@firstoftwo
 \else \expandafter \@secondoftwo
 \fi
}%
\providecommand \@ifx [1]{%
 \ifx #1\expandafter \@firstoftwo
 \else \expandafter \@secondoftwo
 \fi
}%
\providecommand \natexlab [1]{#1}%
\providecommand \enquote  [1]{``#1''}%
\providecommand \bibnamefont  [1]{#1}%
\providecommand \bibfnamefont [1]{#1}%
\providecommand \citenamefont [1]{#1}%
\providecommand \href@noop [0]{\@secondoftwo}%
\providecommand \href [0]{\begingroup \@sanitize@url \@href}%
\providecommand \@href[1]{\@@startlink{#1}\@@href}%
\providecommand \@@href[1]{\endgroup#1\@@endlink}%
\providecommand \@sanitize@url [0]{\catcode `\\12\catcode `\$12\catcode
  `\&12\catcode `\#12\catcode `\^12\catcode `\_12\catcode `\%12\relax}%
\providecommand \@@startlink[1]{}%
\providecommand \@@endlink[0]{}%
\providecommand \url  [0]{\begingroup\@sanitize@url \@url }%
\providecommand \@url [1]{\endgroup\@href {#1}{\urlprefix }}%
\providecommand \urlprefix  [0]{URL }%
\providecommand \Eprint [0]{\href }%
\providecommand \doibase [0]{http://dx.doi.org/}%
\providecommand \selectlanguage [0]{\@gobble}%
\providecommand \bibinfo  [0]{\@secondoftwo}%
\providecommand \bibfield  [0]{\@secondoftwo}%
\providecommand \translation [1]{[#1]}%
\providecommand \BibitemOpen [0]{}%
\providecommand \bibitemStop [0]{}%
\providecommand \bibitemNoStop [0]{.\EOS\space}%
\providecommand \EOS [0]{\spacefactor3000\relax}%
\providecommand \BibitemShut  [1]{\csname bibitem#1\endcsname}%
\let\auto@bib@innerbib\@empty
\bibitem [{\citenamefont {Heshami}\ \emph {et~al.}(2016)\citenamefont
  {Heshami}, \citenamefont {England}, \citenamefont {Humphreys}, \citenamefont
  {Bustard}, \citenamefont {Acosta}, \citenamefont {Nunn},\ and\ \citenamefont
  {Sussman}}]{heshami2016quantum}%
  \BibitemOpen
  \bibfield  {author} {\bibinfo {author} {\bibfnamefont {K.}~\bibnamefont
  {Heshami}}, \bibinfo {author} {\bibfnamefont {D.~G.}\ \bibnamefont
  {England}}, \bibinfo {author} {\bibfnamefont {P.~C.}\ \bibnamefont
  {Humphreys}}, \bibinfo {author} {\bibfnamefont {P.~J.}\ \bibnamefont
  {Bustard}}, \bibinfo {author} {\bibfnamefont {V.~M.}\ \bibnamefont {Acosta}},
  \bibinfo {author} {\bibfnamefont {J.}~\bibnamefont {Nunn}}, \ and\ \bibinfo
  {author} {\bibfnamefont {B.~J.}\ \bibnamefont {Sussman}},\ }\href
  {https://www.tandfonline.com/doi/full/10.1080/09500340.2016.1148212}
  {\bibfield  {journal} {\bibinfo  {journal} {Journal of modern optics}\
  }\textbf {\bibinfo {volume} {63}},\ \bibinfo {pages} {2005} (\bibinfo {year}
  {2016})}\BibitemShut {NoStop}%
\bibitem [{\citenamefont {Bussieres}\ \emph {et~al.}(2013)\citenamefont
  {Bussieres}, \citenamefont {Sangouard}, \citenamefont {Afzelius},
  \citenamefont {De~Riedmatten}, \citenamefont {Simon},\ and\ \citenamefont
  {Tittel}}]{bussieres2013prospective}%
  \BibitemOpen
  \bibfield  {author} {\bibinfo {author} {\bibfnamefont {F.}~\bibnamefont
  {Bussieres}}, \bibinfo {author} {\bibfnamefont {N.}~\bibnamefont
  {Sangouard}}, \bibinfo {author} {\bibfnamefont {M.}~\bibnamefont {Afzelius}},
  \bibinfo {author} {\bibfnamefont {H.}~\bibnamefont {De~Riedmatten}}, \bibinfo
  {author} {\bibfnamefont {C.}~\bibnamefont {Simon}}, \ and\ \bibinfo {author}
  {\bibfnamefont {W.}~\bibnamefont {Tittel}},\ }\href
  {https://www.tandfonline.com/doi/abs/10.1080/09500340.2013.856482} {\bibfield
   {journal} {\bibinfo  {journal} {Journal of Modern Optics}\ }\textbf
  {\bibinfo {volume} {60}},\ \bibinfo {pages} {1519} (\bibinfo {year}
  {2013})}\BibitemShut {NoStop}%
\bibitem [{\citenamefont {Lvovsky}\ \emph {et~al.}(2009)\citenamefont
  {Lvovsky}, \citenamefont {Sanders},\ and\ \citenamefont
  {Tittel}}]{lvovsky2009optical}%
  \BibitemOpen
  \bibfield  {author} {\bibinfo {author} {\bibfnamefont {A.~I.}\ \bibnamefont
  {Lvovsky}}, \bibinfo {author} {\bibfnamefont {B.~C.}\ \bibnamefont
  {Sanders}}, \ and\ \bibinfo {author} {\bibfnamefont {W.}~\bibnamefont
  {Tittel}},\ }\href {https://www.nature.com/articles/nphoton.2009.231}
  {\bibfield  {journal} {\bibinfo  {journal} {Nature photonics}\ }\textbf
  {\bibinfo {volume} {3}},\ \bibinfo {pages} {706} (\bibinfo {year}
  {2009})}\BibitemShut {NoStop}%
\bibitem [{\citenamefont {Kimble}(2008)}]{kimble2008quantum}%
  \BibitemOpen
  \bibfield  {author} {\bibinfo {author} {\bibfnamefont {H.~J.}\ \bibnamefont
  {Kimble}},\ }\href {https://www.nature.com/articles/nature07127} {\bibfield
  {journal} {\bibinfo  {journal} {Nature}\ }\textbf {\bibinfo {volume} {453}},\
  \bibinfo {pages} {1023} (\bibinfo {year} {2008})}\BibitemShut {NoStop}%
\bibitem [{\citenamefont {Simon}(2017)}]{simon2017towards}%
  \BibitemOpen
  \bibfield  {author} {\bibinfo {author} {\bibfnamefont {C.}~\bibnamefont
  {Simon}},\ }\href {https://www.nature.com/articles/s41566-017-0032-0}
  {\bibfield  {journal} {\bibinfo  {journal} {Nature Photonics}\ }\textbf
  {\bibinfo {volume} {11}},\ \bibinfo {pages} {678} (\bibinfo {year}
  {2017})}\BibitemShut {NoStop}%
\bibitem [{\citenamefont {Jennewein}\ \emph {et~al.}(2023)\citenamefont
  {Jennewein}, \citenamefont {Simon}, \citenamefont {Fougeres}, \citenamefont
  {Babin}, \citenamefont {Asadi}, \citenamefont {Kuntz}, \citenamefont
  {Maisonneuve}, \citenamefont {Moffat}, \citenamefont {Mohammadi},\ and\
  \citenamefont {Panneton}}]{jennewein2023qeyssat}%
  \BibitemOpen
  \bibfield  {author} {\bibinfo {author} {\bibfnamefont {T.}~\bibnamefont
  {Jennewein}}, \bibinfo {author} {\bibfnamefont {C.}~\bibnamefont {Simon}},
  \bibinfo {author} {\bibfnamefont {A.}~\bibnamefont {Fougeres}}, \bibinfo
  {author} {\bibfnamefont {F.}~\bibnamefont {Babin}}, \bibinfo {author}
  {\bibfnamefont {F.~K.}\ \bibnamefont {Asadi}}, \bibinfo {author}
  {\bibfnamefont {K.~B.}\ \bibnamefont {Kuntz}}, \bibinfo {author}
  {\bibfnamefont {M.}~\bibnamefont {Maisonneuve}}, \bibinfo {author}
  {\bibfnamefont {B.}~\bibnamefont {Moffat}}, \bibinfo {author} {\bibfnamefont
  {K.}~\bibnamefont {Mohammadi}}, \ and\ \bibinfo {author} {\bibfnamefont
  {D.}~\bibnamefont {Panneton}},\ }\href {https://arxiv.org/abs/2306.02481}
  {\bibfield  {journal} {\bibinfo  {journal} {arXiv preprint arXiv:2306.02481}\
  } (\bibinfo {year} {2023})}\BibitemShut {NoStop}%
\bibitem [{\citenamefont {Phillips}\ \emph {et~al.}(2001)\citenamefont
  {Phillips}, \citenamefont {Fleischhauer}, \citenamefont {Mair}, \citenamefont
  {Walsworth},\ and\ \citenamefont {Lukin}}]{phillips2001storage}%
  \BibitemOpen
  \bibfield  {author} {\bibinfo {author} {\bibfnamefont {D.~F.}\ \bibnamefont
  {Phillips}}, \bibinfo {author} {\bibfnamefont {A.}~\bibnamefont
  {Fleischhauer}}, \bibinfo {author} {\bibfnamefont {A.}~\bibnamefont {Mair}},
  \bibinfo {author} {\bibfnamefont {R.~L.}\ \bibnamefont {Walsworth}}, \ and\
  \bibinfo {author} {\bibfnamefont {M.~D.}\ \bibnamefont {Lukin}},\ }\href
  {https://journals.aps.org/prl/abstract/10.1103/PhysRevLett.86.783} {\bibfield
   {journal} {\bibinfo  {journal} {Physical review letters}\ }\textbf {\bibinfo
  {volume} {86}},\ \bibinfo {pages} {783} (\bibinfo {year} {2001})}\BibitemShut
  {NoStop}%
\bibitem [{\citenamefont {Fleischhauer}\ \emph {et~al.}(2005)\citenamefont
  {Fleischhauer}, \citenamefont {Imamoglu},\ and\ \citenamefont
  {Marangos}}]{fleischhauer2005electromagnetically}%
  \BibitemOpen
  \bibfield  {author} {\bibinfo {author} {\bibfnamefont {M.}~\bibnamefont
  {Fleischhauer}}, \bibinfo {author} {\bibfnamefont {A.}~\bibnamefont
  {Imamoglu}}, \ and\ \bibinfo {author} {\bibfnamefont {J.~P.}\ \bibnamefont
  {Marangos}},\ }\href
  {https://journals.aps.org/rmp/abstract/10.1103/RevModPhys.77.633} {\bibfield
  {journal} {\bibinfo  {journal} {Reviews of modern physics}\ }\textbf
  {\bibinfo {volume} {77}},\ \bibinfo {pages} {633} (\bibinfo {year}
  {2005})}\BibitemShut {NoStop}%
\bibitem [{\citenamefont {Chaneli{\`e}re}\ \emph {et~al.}(2005)\citenamefont
  {Chaneli{\`e}re}, \citenamefont {Matsukevich}, \citenamefont {Jenkins},
  \citenamefont {Lan}, \citenamefont {Kennedy},\ and\ \citenamefont
  {Kuzmich}}]{chaneliere2005storage}%
  \BibitemOpen
  \bibfield  {author} {\bibinfo {author} {\bibfnamefont {T.}~\bibnamefont
  {Chaneli{\`e}re}}, \bibinfo {author} {\bibfnamefont {D.}~\bibnamefont
  {Matsukevich}}, \bibinfo {author} {\bibfnamefont {S.}~\bibnamefont
  {Jenkins}}, \bibinfo {author} {\bibfnamefont {S.-Y.}\ \bibnamefont {Lan}},
  \bibinfo {author} {\bibfnamefont {T.}~\bibnamefont {Kennedy}}, \ and\
  \bibinfo {author} {\bibfnamefont {A.}~\bibnamefont {Kuzmich}},\ }\href
  {https://www.nature.com/articles/nature04315} {\bibfield  {journal} {\bibinfo
   {journal} {Nature}\ }\textbf {\bibinfo {volume} {438}},\ \bibinfo {pages}
  {833} (\bibinfo {year} {2005})}\BibitemShut {NoStop}%
\bibitem [{\citenamefont {Longdell}\ \emph {et~al.}(2005)\citenamefont
  {Longdell}, \citenamefont {Fraval}, \citenamefont {Sellars},\ and\
  \citenamefont {Manson}}]{longdell2005stopped}%
  \BibitemOpen
  \bibfield  {author} {\bibinfo {author} {\bibfnamefont {J.~J.}\ \bibnamefont
  {Longdell}}, \bibinfo {author} {\bibfnamefont {E.}~\bibnamefont {Fraval}},
  \bibinfo {author} {\bibfnamefont {M.~J.}\ \bibnamefont {Sellars}}, \ and\
  \bibinfo {author} {\bibfnamefont {N.~B.}\ \bibnamefont {Manson}},\ }\href
  {https://journals.aps.org/prl/abstract/10.1103/PhysRevLett.95.063601}
  {\bibfield  {journal} {\bibinfo  {journal} {Physical review letters}\
  }\textbf {\bibinfo {volume} {95}},\ \bibinfo {pages} {063601} (\bibinfo
  {year} {2005})}\BibitemShut {NoStop}%
\bibitem [{\citenamefont {Michelberger}\ \emph {et~al.}(2015)\citenamefont
  {Michelberger}, \citenamefont {Champion}, \citenamefont {Sprague},
  \citenamefont {Kaczmarek}, \citenamefont {Barbieri}, \citenamefont {Jin},
  \citenamefont {England}, \citenamefont {Kolthammer}, \citenamefont
  {Saunders}, \citenamefont {Nunn} \emph
  {et~al.}}]{michelberger2015interfacing}%
  \BibitemOpen
  \bibfield  {author} {\bibinfo {author} {\bibfnamefont {P.}~\bibnamefont
  {Michelberger}}, \bibinfo {author} {\bibfnamefont {T.}~\bibnamefont
  {Champion}}, \bibinfo {author} {\bibfnamefont {M.}~\bibnamefont {Sprague}},
  \bibinfo {author} {\bibfnamefont {K.}~\bibnamefont {Kaczmarek}}, \bibinfo
  {author} {\bibfnamefont {M.}~\bibnamefont {Barbieri}}, \bibinfo {author}
  {\bibfnamefont {X.}~\bibnamefont {Jin}}, \bibinfo {author} {\bibfnamefont
  {D.}~\bibnamefont {England}}, \bibinfo {author} {\bibfnamefont
  {W.}~\bibnamefont {Kolthammer}}, \bibinfo {author} {\bibfnamefont
  {D.}~\bibnamefont {Saunders}}, \bibinfo {author} {\bibfnamefont
  {J.}~\bibnamefont {Nunn}},  \emph {et~al.},\ }\href
  {https://iopscience.iop.org/article/10.1088/1367-2630/17/4/043006/meta}
  {\bibfield  {journal} {\bibinfo  {journal} {New Journal of Physics}\ }\textbf
  {\bibinfo {volume} {17}},\ \bibinfo {pages} {043006} (\bibinfo {year}
  {2015})}\BibitemShut {NoStop}%
\bibitem [{\citenamefont {Saunders}\ \emph {et~al.}(2016)\citenamefont
  {Saunders}, \citenamefont {Munns}, \citenamefont {Champion}, \citenamefont
  {Qiu}, \citenamefont {Kaczmarek}, \citenamefont {Poem}, \citenamefont
  {Ledingham}, \citenamefont {Walmsley},\ and\ \citenamefont
  {Nunn}}]{saunders2016cavity}%
  \BibitemOpen
  \bibfield  {author} {\bibinfo {author} {\bibfnamefont {D.}~\bibnamefont
  {Saunders}}, \bibinfo {author} {\bibfnamefont {J.}~\bibnamefont {Munns}},
  \bibinfo {author} {\bibfnamefont {T.}~\bibnamefont {Champion}}, \bibinfo
  {author} {\bibfnamefont {C.}~\bibnamefont {Qiu}}, \bibinfo {author}
  {\bibfnamefont {K.}~\bibnamefont {Kaczmarek}}, \bibinfo {author}
  {\bibfnamefont {E.}~\bibnamefont {Poem}}, \bibinfo {author} {\bibfnamefont
  {P.}~\bibnamefont {Ledingham}}, \bibinfo {author} {\bibfnamefont
  {I.}~\bibnamefont {Walmsley}}, \ and\ \bibinfo {author} {\bibfnamefont
  {J.}~\bibnamefont {Nunn}},\ }\href
  {https://journals.aps.org/prl/abstract/10.1103/PhysRevLett.116.090501}
  {\bibfield  {journal} {\bibinfo  {journal} {Physical review letters}\
  }\textbf {\bibinfo {volume} {116}},\ \bibinfo {pages} {090501} (\bibinfo
  {year} {2016})}\BibitemShut {NoStop}%
\bibitem [{\citenamefont {Guo}\ \emph {et~al.}(2019)\citenamefont {Guo},
  \citenamefont {Feng}, \citenamefont {Yang}, \citenamefont {Yu}, \citenamefont
  {Chen}, \citenamefont {Yuan},\ and\ \citenamefont {Zhang}}]{guo2019high}%
  \BibitemOpen
  \bibfield  {author} {\bibinfo {author} {\bibfnamefont {J.}~\bibnamefont
  {Guo}}, \bibinfo {author} {\bibfnamefont {X.}~\bibnamefont {Feng}}, \bibinfo
  {author} {\bibfnamefont {P.}~\bibnamefont {Yang}}, \bibinfo {author}
  {\bibfnamefont {Z.}~\bibnamefont {Yu}}, \bibinfo {author} {\bibfnamefont
  {L.}~\bibnamefont {Chen}}, \bibinfo {author} {\bibfnamefont {C.-H.}\
  \bibnamefont {Yuan}}, \ and\ \bibinfo {author} {\bibfnamefont
  {W.}~\bibnamefont {Zhang}},\ }\href
  {https://www.nature.com/articles/s41467-018-08118-5} {\bibfield  {journal}
  {\bibinfo  {journal} {Nature communications}\ }\textbf {\bibinfo {volume}
  {10}},\ \bibinfo {pages} {148} (\bibinfo {year} {2019})}\BibitemShut
  {NoStop}%
\bibitem [{\citenamefont {Heshami}\ \emph {et~al.}(2014)\citenamefont
  {Heshami}, \citenamefont {Santori}, \citenamefont {Khanaliloo}, \citenamefont
  {Healey}, \citenamefont {Acosta}, \citenamefont {Barclay},\ and\
  \citenamefont {Simon}}]{heshami2014raman}%
  \BibitemOpen
  \bibfield  {author} {\bibinfo {author} {\bibfnamefont {K.}~\bibnamefont
  {Heshami}}, \bibinfo {author} {\bibfnamefont {C.}~\bibnamefont {Santori}},
  \bibinfo {author} {\bibfnamefont {B.}~\bibnamefont {Khanaliloo}}, \bibinfo
  {author} {\bibfnamefont {C.}~\bibnamefont {Healey}}, \bibinfo {author}
  {\bibfnamefont {V.~M.}\ \bibnamefont {Acosta}}, \bibinfo {author}
  {\bibfnamefont {P.~E.}\ \bibnamefont {Barclay}}, \ and\ \bibinfo {author}
  {\bibfnamefont {C.}~\bibnamefont {Simon}},\ }\href
  {https://journals.aps.org/pra/abstract/10.1103/PhysRevA.89.040301} {\bibfield
   {journal} {\bibinfo  {journal} {Physical Review A}\ }\textbf {\bibinfo
  {volume} {89}},\ \bibinfo {pages} {040301} (\bibinfo {year}
  {2014})}\BibitemShut {NoStop}%
\bibitem [{\citenamefont {Saglamyurek}\ \emph {et~al.}(2018)\citenamefont
  {Saglamyurek}, \citenamefont {Hrushevskyi}, \citenamefont {Rastogi},
  \citenamefont {Heshami},\ and\ \citenamefont
  {LeBlanc}}]{saglamyurek2018coherent}%
  \BibitemOpen
  \bibfield  {author} {\bibinfo {author} {\bibfnamefont {E.}~\bibnamefont
  {Saglamyurek}}, \bibinfo {author} {\bibfnamefont {T.}~\bibnamefont
  {Hrushevskyi}}, \bibinfo {author} {\bibfnamefont {A.}~\bibnamefont
  {Rastogi}}, \bibinfo {author} {\bibfnamefont {K.}~\bibnamefont {Heshami}}, \
  and\ \bibinfo {author} {\bibfnamefont {L.~J.}\ \bibnamefont {LeBlanc}},\
  }\href {https://www.nature.com/articles/s41566-018-0279-0} {\bibfield
  {journal} {\bibinfo  {journal} {Nature Photonics}\ }\textbf {\bibinfo
  {volume} {12}},\ \bibinfo {pages} {774} (\bibinfo {year} {2018})}\BibitemShut
  {NoStop}%
\bibitem [{\citenamefont {Higginbottom}\ \emph {et~al.}(2023)\citenamefont
  {Higginbottom}, \citenamefont {Asadi}, \citenamefont {Chartrand},
  \citenamefont {Ji}, \citenamefont {Bergeron}, \citenamefont {Thewalt},
  \citenamefont {Simon},\ and\ \citenamefont
  {Simmons}}]{higginbottom2023memory}%
  \BibitemOpen
  \bibfield  {author} {\bibinfo {author} {\bibfnamefont {D.~B.}\ \bibnamefont
  {Higginbottom}}, \bibinfo {author} {\bibfnamefont {F.~K.}\ \bibnamefont
  {Asadi}}, \bibinfo {author} {\bibfnamefont {C.}~\bibnamefont {Chartrand}},
  \bibinfo {author} {\bibfnamefont {J.-W.}\ \bibnamefont {Ji}}, \bibinfo
  {author} {\bibfnamefont {L.}~\bibnamefont {Bergeron}}, \bibinfo {author}
  {\bibfnamefont {M.~L.}\ \bibnamefont {Thewalt}}, \bibinfo {author}
  {\bibfnamefont {C.}~\bibnamefont {Simon}}, \ and\ \bibinfo {author}
  {\bibfnamefont {S.}~\bibnamefont {Simmons}},\ }\href
  {https://journals.aps.org/prxquantum/abstract/10.1103/PRXQuantum.4.020308}
  {\bibfield  {journal} {\bibinfo  {journal} {PRX Quantum}\ }\textbf {\bibinfo
  {volume} {4}},\ \bibinfo {pages} {020308} (\bibinfo {year}
  {2023})}\BibitemShut {NoStop}%
\bibitem [{\citenamefont {Saglamyurek}\ \emph {et~al.}(2019)\citenamefont
  {Saglamyurek}, \citenamefont {Hrushevskyi}, \citenamefont {Cooke},
  \citenamefont {Rastogi},\ and\ \citenamefont
  {LeBlanc}}]{saglamyurek2019single}%
  \BibitemOpen
  \bibfield  {author} {\bibinfo {author} {\bibfnamefont {E.}~\bibnamefont
  {Saglamyurek}}, \bibinfo {author} {\bibfnamefont {T.}~\bibnamefont
  {Hrushevskyi}}, \bibinfo {author} {\bibfnamefont {L.}~\bibnamefont {Cooke}},
  \bibinfo {author} {\bibfnamefont {A.}~\bibnamefont {Rastogi}}, \ and\
  \bibinfo {author} {\bibfnamefont {L.~J.}\ \bibnamefont {LeBlanc}},\ }\href
  {https://journals.aps.org/prresearch/abstract/10.1103/PhysRevResearch.1.022004}
  {\bibfield  {journal} {\bibinfo  {journal} {Physical Review Research}\
  }\textbf {\bibinfo {volume} {1}},\ \bibinfo {pages} {022004} (\bibinfo {year}
  {2019})}\BibitemShut {NoStop}%
\bibitem [{\citenamefont {Lei}\ \emph {et~al.}(2023)\citenamefont {Lei},
  \citenamefont {Kimiaee~Asadi}, \citenamefont {Zhong}, \citenamefont
  {Kuzmich}, \citenamefont {Simon},\ and\ \citenamefont
  {Hosseini}}]{lei2023quantum}%
  \BibitemOpen
  \bibfield  {author} {\bibinfo {author} {\bibfnamefont {Y.}~\bibnamefont
  {Lei}}, \bibinfo {author} {\bibfnamefont {F.}~\bibnamefont {Kimiaee~Asadi}},
  \bibinfo {author} {\bibfnamefont {T.}~\bibnamefont {Zhong}}, \bibinfo
  {author} {\bibfnamefont {A.}~\bibnamefont {Kuzmich}}, \bibinfo {author}
  {\bibfnamefont {C.}~\bibnamefont {Simon}}, \ and\ \bibinfo {author}
  {\bibfnamefont {M.}~\bibnamefont {Hosseini}},\ }\href
  {https://opg.optica.org/abstract.cfm?uri=optica-10-11-1511} {\bibfield
  {journal} {\bibinfo  {journal} {Optica}\ }\textbf {\bibinfo {volume} {10}},\
  \bibinfo {pages} {1511} (\bibinfo {year} {2023})}\BibitemShut {NoStop}%
\bibitem [{\citenamefont {Simon}\ \emph {et~al.}(2010)\citenamefont {Simon},
  \citenamefont {Afzelius}, \citenamefont {Appel}, \citenamefont {Boyer De
  La~Giroday}, \citenamefont {Dewhurst}, \citenamefont {Gisin}, \citenamefont
  {Hu}, \citenamefont {Jelezko}, \citenamefont {Kr{\"o}ll}, \citenamefont
  {M{\"u}ller} \emph {et~al.}}]{simon2010quantum}%
  \BibitemOpen
  \bibfield  {author} {\bibinfo {author} {\bibfnamefont {C.}~\bibnamefont
  {Simon}}, \bibinfo {author} {\bibfnamefont {M.}~\bibnamefont {Afzelius}},
  \bibinfo {author} {\bibfnamefont {J.}~\bibnamefont {Appel}}, \bibinfo
  {author} {\bibfnamefont {A.}~\bibnamefont {Boyer De La~Giroday}}, \bibinfo
  {author} {\bibfnamefont {S.}~\bibnamefont {Dewhurst}}, \bibinfo {author}
  {\bibfnamefont {N.}~\bibnamefont {Gisin}}, \bibinfo {author} {\bibfnamefont
  {C.}~\bibnamefont {Hu}}, \bibinfo {author} {\bibfnamefont {F.}~\bibnamefont
  {Jelezko}}, \bibinfo {author} {\bibfnamefont {S.}~\bibnamefont {Kr{\"o}ll}},
  \bibinfo {author} {\bibfnamefont {J.}~\bibnamefont {M{\"u}ller}},  \emph
  {et~al.},\ }\href
  {https://idp.springer.com/authorize/casa?redirect_uri=https://link.springer.com/article/10.1140/epjd/e2010-00103-y&casa_token=3MxlyXDwjsUAAAAA:1djawvKC2HGzlKtpjPeEL5H7S1vdSVWdnVNR5ffrr9E1rE7w0OsCt7JMcXeDTR1wRlAuRk7AAqy0XWsg}
  {\bibfield  {journal} {\bibinfo  {journal} {The European Physical Journal D}\
  }\textbf {\bibinfo {volume} {58}},\ \bibinfo {pages} {1} (\bibinfo {year}
  {2010})}\BibitemShut {NoStop}%
\bibitem [{\citenamefont {Reim}\ \emph {et~al.}(2010)\citenamefont {Reim},
  \citenamefont {Nunn}, \citenamefont {Lorenz}, \citenamefont {Sussman},
  \citenamefont {Lee}, \citenamefont {Langford}, \citenamefont {Jaksch},\ and\
  \citenamefont {Walmsley}}]{reim2010towards}%
  \BibitemOpen
  \bibfield  {author} {\bibinfo {author} {\bibfnamefont {K.}~\bibnamefont
  {Reim}}, \bibinfo {author} {\bibfnamefont {J.}~\bibnamefont {Nunn}}, \bibinfo
  {author} {\bibfnamefont {V.}~\bibnamefont {Lorenz}}, \bibinfo {author}
  {\bibfnamefont {B.}~\bibnamefont {Sussman}}, \bibinfo {author} {\bibfnamefont
  {K.}~\bibnamefont {Lee}}, \bibinfo {author} {\bibfnamefont {N.}~\bibnamefont
  {Langford}}, \bibinfo {author} {\bibfnamefont {D.}~\bibnamefont {Jaksch}}, \
  and\ \bibinfo {author} {\bibfnamefont {I.}~\bibnamefont {Walmsley}},\ }\href
  {https://www.nature.com/articles/nphoton.2010.30} {\bibfield  {journal}
  {\bibinfo  {journal} {Nature Photonics}\ }\textbf {\bibinfo {volume} {4}},\
  \bibinfo {pages} {218} (\bibinfo {year} {2010})}\BibitemShut {NoStop}%
\bibitem [{\citenamefont {Manz}\ \emph {et~al.}(2007)\citenamefont {Manz},
  \citenamefont {Fernholz}, \citenamefont {Schmiedmayer},\ and\ \citenamefont
  {Pan}}]{manz2007collisional}%
  \BibitemOpen
  \bibfield  {author} {\bibinfo {author} {\bibfnamefont {S.}~\bibnamefont
  {Manz}}, \bibinfo {author} {\bibfnamefont {T.}~\bibnamefont {Fernholz}},
  \bibinfo {author} {\bibfnamefont {J.}~\bibnamefont {Schmiedmayer}}, \ and\
  \bibinfo {author} {\bibfnamefont {J.-W.}\ \bibnamefont {Pan}},\ }\href
  {https://journals.aps.org/pra/abstract/10.1103/PhysRevA.75.040101} {\bibfield
   {journal} {\bibinfo  {journal} {Physical Review A—Atomic, Molecular, and
  Optical Physics}\ }\textbf {\bibinfo {volume} {75}},\ \bibinfo {pages}
  {040101} (\bibinfo {year} {2007})}\BibitemShut {NoStop}%
\bibitem [{\citenamefont {Lauk}\ \emph {et~al.}(2013)\citenamefont {Lauk},
  \citenamefont {O'Brien},\ and\ \citenamefont
  {Fleischhauer}}]{lauk2013fidelity}%
  \BibitemOpen
  \bibfield  {author} {\bibinfo {author} {\bibfnamefont {N.}~\bibnamefont
  {Lauk}}, \bibinfo {author} {\bibfnamefont {C.}~\bibnamefont {O'Brien}}, \
  and\ \bibinfo {author} {\bibfnamefont {M.}~\bibnamefont {Fleischhauer}},\
  }\href {https://journals.aps.org/pra/abstract/10.1103/PhysRevA.88.013823}
  {\bibfield  {journal} {\bibinfo  {journal} {Physical Review A—Atomic,
  Molecular, and Optical Physics}\ }\textbf {\bibinfo {volume} {88}},\ \bibinfo
  {pages} {013823} (\bibinfo {year} {2013})}\BibitemShut {NoStop}%
\bibitem [{\citenamefont {Gorshkov}\ \emph {et~al.}(2007)\citenamefont
  {Gorshkov}, \citenamefont {Andr{\'e}}, \citenamefont {Lukin},\ and\
  \citenamefont {S{\o}rensen}}]{gorshkov2007photon}%
  \BibitemOpen
  \bibfield  {author} {\bibinfo {author} {\bibfnamefont {A.~V.}\ \bibnamefont
  {Gorshkov}}, \bibinfo {author} {\bibfnamefont {A.}~\bibnamefont {Andr{\'e}}},
  \bibinfo {author} {\bibfnamefont {M.~D.}\ \bibnamefont {Lukin}}, \ and\
  \bibinfo {author} {\bibfnamefont {A.~S.}\ \bibnamefont {S{\o}rensen}},\
  }\href {https://journals.aps.org/pra/abstract/10.1103/PhysRevA.76.033804}
  {\bibfield  {journal} {\bibinfo  {journal} {Physical Review A—Atomic,
  Molecular, and Optical Physics}\ }\textbf {\bibinfo {volume} {76}},\ \bibinfo
  {pages} {033804} (\bibinfo {year} {2007})}\BibitemShut {NoStop}%
\bibitem [{\citenamefont {Fleischhauer}\ and\ \citenamefont
  {Lukin}(2002)}]{fleischhauer2002quantum}%
  \BibitemOpen
  \bibfield  {author} {\bibinfo {author} {\bibfnamefont {M.}~\bibnamefont
  {Fleischhauer}}\ and\ \bibinfo {author} {\bibfnamefont {M.~D.}\ \bibnamefont
  {Lukin}},\ }\href
  {https://journals.aps.org/pra/abstract/10.1103/PhysRevA.65.022314} {\bibfield
   {journal} {\bibinfo  {journal} {Physical Review A}\ }\textbf {\bibinfo
  {volume} {65}},\ \bibinfo {pages} {022314} (\bibinfo {year}
  {2002})}\BibitemShut {NoStop}%
\bibitem [{\citenamefont {Clerk}\ \emph {et~al.}(2010)\citenamefont {Clerk},
  \citenamefont {Devoret}, \citenamefont {Girvin}, \citenamefont {Marquardt},\
  and\ \citenamefont {Schoelkopf}}]{clerk2010introduction}%
  \BibitemOpen
  \bibfield  {author} {\bibinfo {author} {\bibfnamefont {A.~A.}\ \bibnamefont
  {Clerk}}, \bibinfo {author} {\bibfnamefont {M.~H.}\ \bibnamefont {Devoret}},
  \bibinfo {author} {\bibfnamefont {S.~M.}\ \bibnamefont {Girvin}}, \bibinfo
  {author} {\bibfnamefont {F.}~\bibnamefont {Marquardt}}, \ and\ \bibinfo
  {author} {\bibfnamefont {R.~J.}\ \bibnamefont {Schoelkopf}},\ }\href
  {https://journals.aps.org/rmp/abstract/10.1103/RevModPhys.82.1155} {\bibfield
   {journal} {\bibinfo  {journal} {Reviews of Modern Physics}\ }\textbf
  {\bibinfo {volume} {82}},\ \bibinfo {pages} {1155} (\bibinfo {year}
  {2010})}\BibitemShut {NoStop}%
\bibitem [{Sup()}]{SupplementaryMaterial}%
  \BibitemOpen
  \href@noop {} {\enquote {\bibinfo {title} {See supplemental material for more
  details},}\ }\BibitemShut {NoStop}%
\bibitem [{\citenamefont {Doherty}\ \emph {et~al.}(2012)\citenamefont
  {Doherty}, \citenamefont {Dolde}, \citenamefont {Fedder}, \citenamefont
  {Jelezko}, \citenamefont {Wrachtrup}, \citenamefont {Manson},\ and\
  \citenamefont {Hollenberg}}]{doherty2012theory}%
  \BibitemOpen
  \bibfield  {author} {\bibinfo {author} {\bibfnamefont {M.}~\bibnamefont
  {Doherty}}, \bibinfo {author} {\bibfnamefont {F.}~\bibnamefont {Dolde}},
  \bibinfo {author} {\bibfnamefont {H.}~\bibnamefont {Fedder}}, \bibinfo
  {author} {\bibfnamefont {F.}~\bibnamefont {Jelezko}}, \bibinfo {author}
  {\bibfnamefont {J.}~\bibnamefont {Wrachtrup}}, \bibinfo {author}
  {\bibfnamefont {N.}~\bibnamefont {Manson}}, \ and\ \bibinfo {author}
  {\bibfnamefont {L.}~\bibnamefont {Hollenberg}},\ }\href
  {https://journals.aps.org/prb/abstract/10.1103/PhysRevB.85.205203} {\bibfield
   {journal} {\bibinfo  {journal} {Physical Review B—Condensed Matter and
  Materials Physics}\ }\textbf {\bibinfo {volume} {85}},\ \bibinfo {pages}
  {205203} (\bibinfo {year} {2012})}\BibitemShut {NoStop}%
\bibitem [{\citenamefont {Doherty}\ \emph {et~al.}(2013)\citenamefont
  {Doherty}, \citenamefont {Manson}, \citenamefont {Delaney}, \citenamefont
  {Jelezko}, \citenamefont {Wrachtrup},\ and\ \citenamefont
  {Hollenberg}}]{doherty2013nitrogen}%
  \BibitemOpen
  \bibfield  {author} {\bibinfo {author} {\bibfnamefont {M.~W.}\ \bibnamefont
  {Doherty}}, \bibinfo {author} {\bibfnamefont {N.~B.}\ \bibnamefont {Manson}},
  \bibinfo {author} {\bibfnamefont {P.}~\bibnamefont {Delaney}}, \bibinfo
  {author} {\bibfnamefont {F.}~\bibnamefont {Jelezko}}, \bibinfo {author}
  {\bibfnamefont {J.}~\bibnamefont {Wrachtrup}}, \ and\ \bibinfo {author}
  {\bibfnamefont {L.~C.}\ \bibnamefont {Hollenberg}},\ }\href
  {https://journals.aps.org/prb/abstract/10.1103/PhysRevB.85.205203} {\bibfield
   {journal} {\bibinfo  {journal} {Physics Reports}\ }\textbf {\bibinfo
  {volume} {528}},\ \bibinfo {pages} {1} (\bibinfo {year} {2013})}\BibitemShut
  {NoStop}%
\bibitem [{\citenamefont {Rastogi}\ \emph {et~al.}(2019)\citenamefont
  {Rastogi}, \citenamefont {Saglamyurek}, \citenamefont {Hrushevskyi},
  \citenamefont {Hubele},\ and\ \citenamefont
  {LeBlanc}}]{rastogi2019discerning}%
  \BibitemOpen
  \bibfield  {author} {\bibinfo {author} {\bibfnamefont {A.}~\bibnamefont
  {Rastogi}}, \bibinfo {author} {\bibfnamefont {E.}~\bibnamefont
  {Saglamyurek}}, \bibinfo {author} {\bibfnamefont {T.}~\bibnamefont
  {Hrushevskyi}}, \bibinfo {author} {\bibfnamefont {S.}~\bibnamefont {Hubele}},
  \ and\ \bibinfo {author} {\bibfnamefont {L.~J.}\ \bibnamefont {LeBlanc}},\
  }\href {https://journals.aps.org/pra/abstract/10.1103/PhysRevA.100.012314}
  {\bibfield  {journal} {\bibinfo  {journal} {Physical Review A}\ }\textbf
  {\bibinfo {volume} {100}},\ \bibinfo {pages} {012314} (\bibinfo {year}
  {2019})}\BibitemShut {NoStop}%
\bibitem [{\citenamefont {Nunn}\ \emph {et~al.}(2017)\citenamefont {Nunn},
  \citenamefont {Munns}, \citenamefont {Thomas}, \citenamefont {Kaczmarek},
  \citenamefont {Qiu}, \citenamefont {Feizpour}, \citenamefont {Poem},
  \citenamefont {Brecht}, \citenamefont {Saunders}, \citenamefont {Ledingham}
  \emph {et~al.}}]{nunn2017theory}%
  \BibitemOpen
  \bibfield  {author} {\bibinfo {author} {\bibfnamefont {J.}~\bibnamefont
  {Nunn}}, \bibinfo {author} {\bibfnamefont {J.}~\bibnamefont {Munns}},
  \bibinfo {author} {\bibfnamefont {S.}~\bibnamefont {Thomas}}, \bibinfo
  {author} {\bibfnamefont {K.~T.}\ \bibnamefont {Kaczmarek}}, \bibinfo {author}
  {\bibfnamefont {C.}~\bibnamefont {Qiu}}, \bibinfo {author} {\bibfnamefont
  {A.}~\bibnamefont {Feizpour}}, \bibinfo {author} {\bibfnamefont
  {E.}~\bibnamefont {Poem}}, \bibinfo {author} {\bibfnamefont {B.}~\bibnamefont
  {Brecht}}, \bibinfo {author} {\bibfnamefont {D.}~\bibnamefont {Saunders}},
  \bibinfo {author} {\bibfnamefont {P.~M.}\ \bibnamefont {Ledingham}},  \emph
  {et~al.},\ }\href
  {https://journals.aps.org/pra/abstract/10.1103/PhysRevA.96.012338} {\bibfield
   {journal} {\bibinfo  {journal} {Physical Review A}\ }\textbf {\bibinfo
  {volume} {96}},\ \bibinfo {pages} {012338} (\bibinfo {year}
  {2017})}\BibitemShut {NoStop}%
\bibitem [{\citenamefont {Thomas}\ \emph {et~al.}(2019)\citenamefont {Thomas},
  \citenamefont {Hird}, \citenamefont {Munns}, \citenamefont {Brecht},
  \citenamefont {Saunders}, \citenamefont {Nunn}, \citenamefont {Walmsley},\
  and\ \citenamefont {Ledingham}}]{thomas2019raman}%
  \BibitemOpen
  \bibfield  {author} {\bibinfo {author} {\bibfnamefont {S.~E.}\ \bibnamefont
  {Thomas}}, \bibinfo {author} {\bibfnamefont {T.~M.}\ \bibnamefont {Hird}},
  \bibinfo {author} {\bibfnamefont {J.~H.}\ \bibnamefont {Munns}}, \bibinfo
  {author} {\bibfnamefont {B.}~\bibnamefont {Brecht}}, \bibinfo {author}
  {\bibfnamefont {D.~J.}\ \bibnamefont {Saunders}}, \bibinfo {author}
  {\bibfnamefont {J.}~\bibnamefont {Nunn}}, \bibinfo {author} {\bibfnamefont
  {I.~A.}\ \bibnamefont {Walmsley}}, \ and\ \bibinfo {author} {\bibfnamefont
  {P.~M.}\ \bibnamefont {Ledingham}},\ }\href
  {https://journals.aps.org/pra/abstract/10.1103/PhysRevA.100.033801}
  {\bibfield  {journal} {\bibinfo  {journal} {Physical Review A}\ }\textbf
  {\bibinfo {volume} {100}},\ \bibinfo {pages} {033801} (\bibinfo {year}
  {2019})}\BibitemShut {NoStop}%
\bibitem [{\citenamefont {Steck}(2001)}]{steck2001rubidium}%
  \BibitemOpen
  \bibfield  {author} {\bibinfo {author} {\bibfnamefont {D.~A.}\ \bibnamefont
  {Steck}},\ }\href {https://www.steck.us/alkalidata/rubidium85numbers.pdf} {\
  (\bibinfo {year} {2001})}\BibitemShut {NoStop}%
\bibitem [{\citenamefont {Ma}\ \emph {et~al.}(2022)\citenamefont {Ma},
  \citenamefont {Lei}, \citenamefont {Yan}, \citenamefont {Li}, \citenamefont
  {Chai}, \citenamefont {Yan}, \citenamefont {Jia}, \citenamefont {Xie},\ and\
  \citenamefont {Peng}}]{ma2022high}%
  \BibitemOpen
  \bibfield  {author} {\bibinfo {author} {\bibfnamefont {L.}~\bibnamefont
  {Ma}}, \bibinfo {author} {\bibfnamefont {X.}~\bibnamefont {Lei}}, \bibinfo
  {author} {\bibfnamefont {J.}~\bibnamefont {Yan}}, \bibinfo {author}
  {\bibfnamefont {R.}~\bibnamefont {Li}}, \bibinfo {author} {\bibfnamefont
  {T.}~\bibnamefont {Chai}}, \bibinfo {author} {\bibfnamefont {Z.}~\bibnamefont
  {Yan}}, \bibinfo {author} {\bibfnamefont {X.}~\bibnamefont {Jia}}, \bibinfo
  {author} {\bibfnamefont {C.}~\bibnamefont {Xie}}, \ and\ \bibinfo {author}
  {\bibfnamefont {K.}~\bibnamefont {Peng}},\ }\href
  {https://www.nature.com/articles/s41467-022-30077-1} {\bibfield  {journal}
  {\bibinfo  {journal} {Nature communications}\ }\textbf {\bibinfo {volume}
  {13}},\ \bibinfo {pages} {2368} (\bibinfo {year} {2022})}\BibitemShut
  {NoStop}%
\bibitem [{\citenamefont {Wu}\ \emph {et~al.}(2025)\citenamefont {Wu},
  \citenamefont {Guo}, \citenamefont {Yu}, \citenamefont {Huang}, \citenamefont
  {Yuan}, \citenamefont {Zhang},\ and\ \citenamefont {Chen}}]{wu2025ai}%
  \BibitemOpen
  \bibfield  {author} {\bibinfo {author} {\bibfnamefont {Z.}~\bibnamefont
  {Wu}}, \bibinfo {author} {\bibfnamefont {J.}~\bibnamefont {Guo}}, \bibinfo
  {author} {\bibfnamefont {Z.}~\bibnamefont {Yu}}, \bibinfo {author}
  {\bibfnamefont {W.}~\bibnamefont {Huang}}, \bibinfo {author} {\bibfnamefont
  {C.-H.}\ \bibnamefont {Yuan}}, \bibinfo {author} {\bibfnamefont
  {W.}~\bibnamefont {Zhang}}, \ and\ \bibinfo {author} {\bibfnamefont
  {L.}~\bibnamefont {Chen}},\ }\href {https://arxiv.org/abs/2503.11098}
  {\bibfield  {journal} {\bibinfo  {journal} {arXiv preprint arXiv:2503.11098}\
  } (\bibinfo {year} {2025})}\BibitemShut {NoStop}%
\bibitem [{\citenamefont {Zhang}\ \emph {et~al.}(2014)\citenamefont {Zhang},
  \citenamefont {Guo}, \citenamefont {Chen}, \citenamefont {Yuan},
  \citenamefont {Ou},\ and\ \citenamefont {Zhang}}]{PhysRevA.90.033823}%
  \BibitemOpen
  \bibfield  {author} {\bibinfo {author} {\bibfnamefont {K.}~\bibnamefont
  {Zhang}}, \bibinfo {author} {\bibfnamefont {J.}~\bibnamefont {Guo}}, \bibinfo
  {author} {\bibfnamefont {L.~Q.}\ \bibnamefont {Chen}}, \bibinfo {author}
  {\bibfnamefont {C.}~\bibnamefont {Yuan}}, \bibinfo {author} {\bibfnamefont
  {Z.~Y.}\ \bibnamefont {Ou}}, \ and\ \bibinfo {author} {\bibfnamefont
  {W.}~\bibnamefont {Zhang}},\ }\href {\doibase 10.1103/PhysRevA.90.033823}
  {\bibfield  {journal} {\bibinfo  {journal} {Phys. Rev. A}\ }\textbf {\bibinfo
  {volume} {90}},\ \bibinfo {pages} {033823} (\bibinfo {year}
  {2014})}\BibitemShut {NoStop}%
\bibitem [{\citenamefont {Hird}(2021)}]{hird2021engineering}%
  \BibitemOpen
  \bibfield  {author} {\bibinfo {author} {\bibfnamefont {T.~M.}\ \bibnamefont
  {Hird}},\ }\emph {\bibinfo {title} {Engineering a noise-free quantum memory
  for temporal mode manipulation}},\ \href
  {https://ora.ox.ac.uk/objects/uuid:f3d40114-a3be-43f9-812c-aec593132b9a}
  {Ph.D. thesis},\ \bibinfo  {school} {University of Oxford} (\bibinfo {year}
  {2021})\BibitemShut {NoStop}%
\end{thebibliography}%

\section*{Supplementary material}\label{sm}

\setcounter{section}{0}
\renewcommand{\thesection}{S\Roman{section}}

\setcounter{equation}{0}
\setcounter{figure}{0}
\setcounter{table}{0}

\renewcommand{\theequation}{S\arabic{equation}}
\renewcommand{\thefigure}{S\arabic{figure}}
\renewcommand{\thetable}{S\arabic{table}}

\subsection{NV center}\label{app:Hamiltonian}

For an ensemble of NV center, the Hamiltonian of the system is given by
$\hat{H} = \hat{H}_{0} +\hat{H}_{\text int}$.
Here $\hat{H}_{0}$ is the free Hamiltonian of the system $
\hat{H}_{0}=\sum_{i=1}^{N}\sum_{j=1}^{9}e^i_j \hat{\sigma}^i_{jj}+\hbar \omega_c \hat{a}^\dagger \hat{a} $
where $j = 1,..., 9$ refers to the 9 energy levels of the system, $e^i_j$ is the eigenenergy of the i-th NV center, $\hat{\sigma}_{jj}=\ket{j}\bra{j}$, $\omega_c$ is the cavity frequency, $\hat{a}$ is the cavity annihilation operator, and $N$ represents the number of centers in the ensemble.  
$\hat{H}_{int}$ is the interaction Hamiltonian between NV-cavity and NV-control field which is given by:
\begin{equation}
\begin{aligned}
-\hat{H}_{\text {int}}/\hbar&=\sum_{i=1}^{N}\sum_{j=1}^{3}\sum_{k=4}^{9}\hat{a}G_{jk}\hat{\sigma}^i_{kj} +\Omega_{jk}\hat{\sigma}^i_{kj}e^{-i \omega_2 t}+\text{H.c}.
\end{aligned}
\end{equation}
Here $j=1,2,3$ refer to the ground states. We then define a set of time-independent collective atomic operators as $\hat{\sigma}^\prime_{kj}=\sum_{i=1}^{N}\hat{\sigma}^i_{kj}$ and
$\hat{\sigma}^\prime_{ll}=\sum_{i=1}^{N}\hat{\sigma}^i_{ll}$ such that ${[\hat{\sigma}_{k k}^{\prime}, \hat{\sigma}_{k^{\prime} j}^{\prime}]=\delta_{k k^{\prime}} \hat{\sigma}_{k j}^{\prime}}$. \begingroup
Using these collective atomic operators, the Hamiltonian of the system in the rotating frame can be written as

\begin{equation}
\begin{aligned}
&\hat{\tilde{H}} / \hbar =\!\sum_{k=4}^{9} \{ {\Delta_k \hat{\sigma}'_{k k}}\!- \!\hat{a} G_{1k} \hat{\sigma}_{k 1}^{\prime} e^{i\omega_{22} t}\!-\Omega_{1k} \hat{\sigma}_{k 1}^{\prime} e^{i\omega_{33} t} \\&-\! \hat{a} G_{2k} \hat{\sigma}_{k 2}^{\prime}-\!\Omega_{2k} \hat{\sigma}_{k 2}^{\prime} e^{-i\delta t} - \!\hat{a} G_{3k} \hat{\sigma}_{k 3}^{\prime} e^{i\delta t}-\!\Omega_{3k} \hat{\sigma}_{k 3}^{\prime} \}\!-\!\!\text { H.c. }\label{H}
\end{aligned}
\end{equation}
where $\delta$ is the splitting between ground states $\ket{2}$ and $\ket{3}$, $\omega_{22}=e_{22} / \hbar$, $\omega_{33}=e_{33} / \hbar$, $e_{jj}$ is the eigenenergy of the system, and $\Delta_k=\omega_{k k}-\omega_c-\omega_{22}=\omega_{k 2}-\omega_c$ is the detunings for
the $k^{th}$ excited states with respect to the ninth level.

Using the a semi-classical approximation the equations of motion can be written as:
\begin{equation}
\begin{aligned}
&\dot{\sigma}^\prime_{23}(t) = -\gamma_s\sigma^\prime_{23} -i\sum_{k=4}^{9} \{ a G_{2k}
\sigma^\prime_{k3}+\Omega_{2k}\sigma^\prime_{k3}e^{-i\delta t}\!-\Omega^\star_{3k}
\sigma^\prime_{2k}\\&-a^\dagger G^\star_{3k}\sigma^\prime_{2k} e^{-i\delta t} \}, 
\\& 
\dot{\sigma}^\prime_{3k}(t) = -
(i\Delta_k+ \gamma_d(T) + \gamma_{e})\sigma^\prime_{3k}+ia G_{2k}
\sigma^\prime_{32}\\&+i\Omega_{2k}
\sigma^\prime_{32}e^{-i\delta t}\!
+ia G_{3k}
(\sigma^\prime_{33}-\sigma^\prime_{kk})e^{i\delta t}+i\Omega_{3k}(\sigma^\prime_{33}-
\sigma^\prime_{kk})\\&
+ia G_{1k}
\sigma^\prime_{31}e^{i\omega_{22} t}+i\Omega_{1k}
\sigma^\prime_{31}e^{i\omega_{33} t},\\
&\dot{\sigma}^\prime_{2k}(t) =-(i\Delta_k+ \gamma_d(T) + \gamma_{e})\sigma^\prime_{2k} +i
a G_{3k}\sigma^\prime_{23}e^{i\delta t}\\&+iaG_{2k}(\sigma^\prime_{22}-\sigma^\prime_{kk})+i\Omega_{2k}(\sigma^\prime_{22}-\sigma^\prime_{kk}) e^{-i\delta t}+i\Omega_{3k}\sigma^\prime_{23}\\&
+i\Omega_{1k}\sigma^\prime_{21}e^{i\omega_{33} t}
+i a G_{1k}\sigma^\prime_{21}e^{i\omega_{22} t},
\end{aligned}
\end{equation} 
where $\gamma_s$ is the spin inhomogeneous broadening, $\gamma_{e}$ is to the optical inhomogeneous broadening, and $\gamma_d(T)$ is the temperature dependent decoherence rate of the optical transitions. For temperatures up to 100 K, the temperature dependency of the latter follows $\gamma_d(T) = \Gamma(T)/2$, where $\Gamma(T)=\gamma_0 + c r T^5$,  $\gamma_0 = 2\pi\times16.2$ MHz, $c = 9.2\times10^{-7}K^{-5}$, and $r = (12.5 \text{ns})^{-1}$ \cite{fu2009observation}.  

In the rotating frame, the evolution of the cavity field is also given by
 \begin{equation}
\begin{aligned}
&\dot{a}=-\kappa a+\sqrt{2 \kappa} a_{\text {in}}+i \sum_{k=4}^9 \{ G^*_{1k}\sigma_{1k}^{\prime} e^{-i\omega_{22} t}+ G^*_{2k}\sigma_{2k}^{\prime} \\&+G^*_{3k}\sigma_{3k}^{\prime} e^{-i\delta t}\}.
\end{aligned}
\end{equation}
 Utilizing this equation alongside the input-output relation $a_{\text{out}}(t) = \sqrt{2\kappa}a(t)-a_{\text{in}}(t)$, one can then determine the output field $a_{\text{out}}$.

Similarly, in a simplified 4-level system discussed in the paper, the equations of motion are given by:
\begin{equation}
\begin{aligned}
&\dot{\sigma}^\prime_{28}(t) =-(i\Delta_8+\gamma_d(T)+\gamma_{e})\sigma^\prime_{28}+i
aG_{38}\sigma^\prime_{23}e^{i\delta t}\\&+i\Omega_{28}N e^{-i\delta t},\\&
\sigma^\prime_{29}(t) =-(\gamma_d(T) +\gamma_{e})\sigma^\prime_{29}+ia G_{29}N+ i\Omega_{39}\sigma^\prime_{23},\\&
\dot{\sigma}^\prime_{23}(t) = -\gamma_s\sigma^\prime_{23} -iaG_{29}
\sigma^\prime_{93}+i\Omega^*_{39}
\sigma^\prime_{29}+ia^\dagger G^*_{38}
\sigma^\prime_{28} e^{-i\delta t},\\&
\dot{\sigma}^\prime_{39}(t)= -
(\gamma_d(T) + \gamma_{e})\sigma^\prime_{39}+ia G_{29}
\sigma^\prime_{32},\\&
\dot{a}=-\kappa a+\sqrt{2 \kappa} a_{\text {in}}+i G^*_{29}\sigma_{29}^{\prime}.
\end{aligned}
\end{equation}

\begin{figure}[b]
    \centering
\includegraphics[scale=0.55]{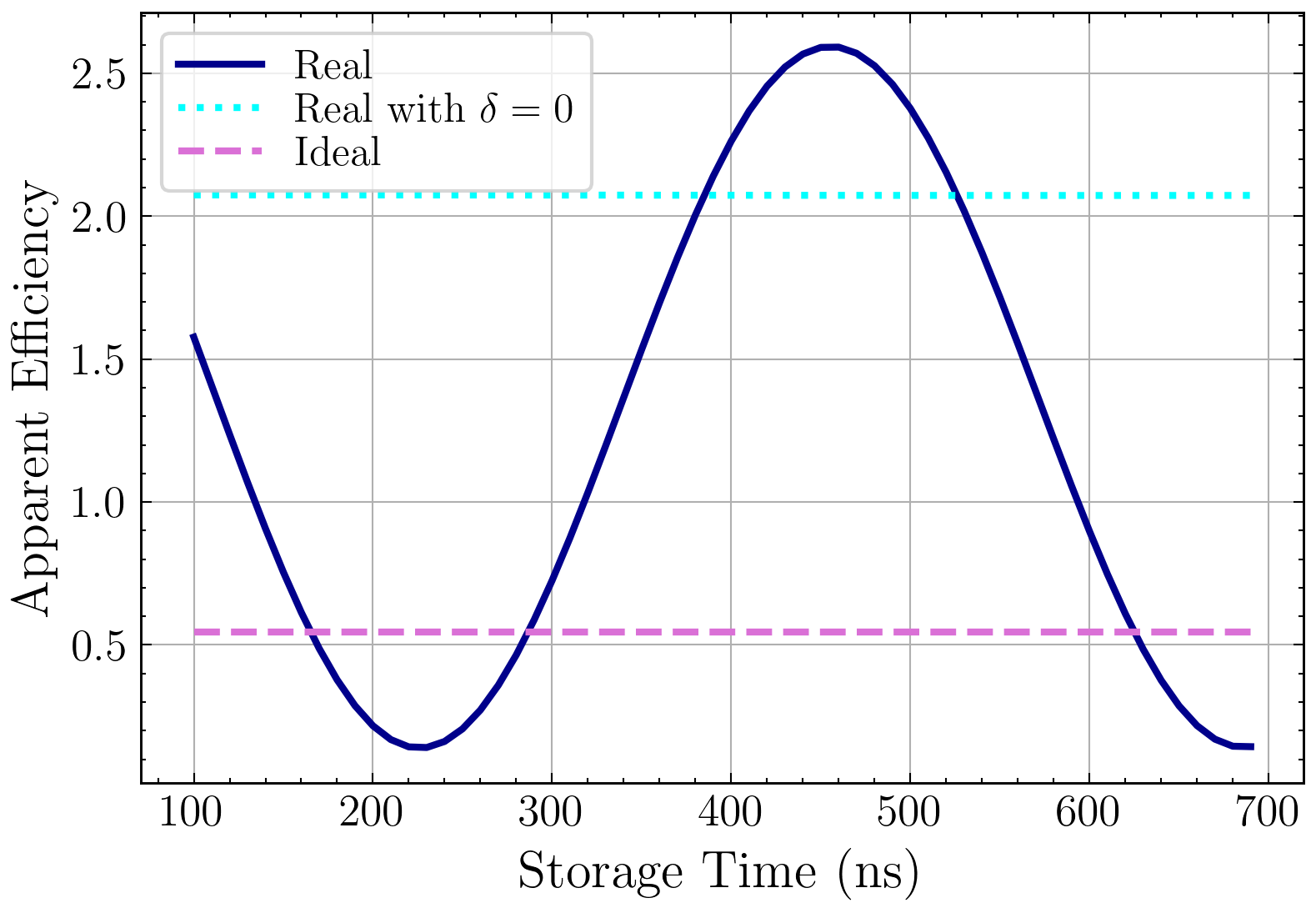}
\caption{Numerical results: Apparent efficiency of the memory  as a function of storage time for $T = 2$ K. The ideal case represents the memory efficiency when all unwanted couplings are eliminated, while the real case accounts for all unwanted couplings present in the 9-level system. The parameters used here are the same as those used to plot Figure 2 of the paper. Here, we assume that spin decoherence and decay rates are negligible.}
\label{oscillation}
\end{figure}
\begin{figure*}
    \centering
{\includegraphics[width=5.8cm]{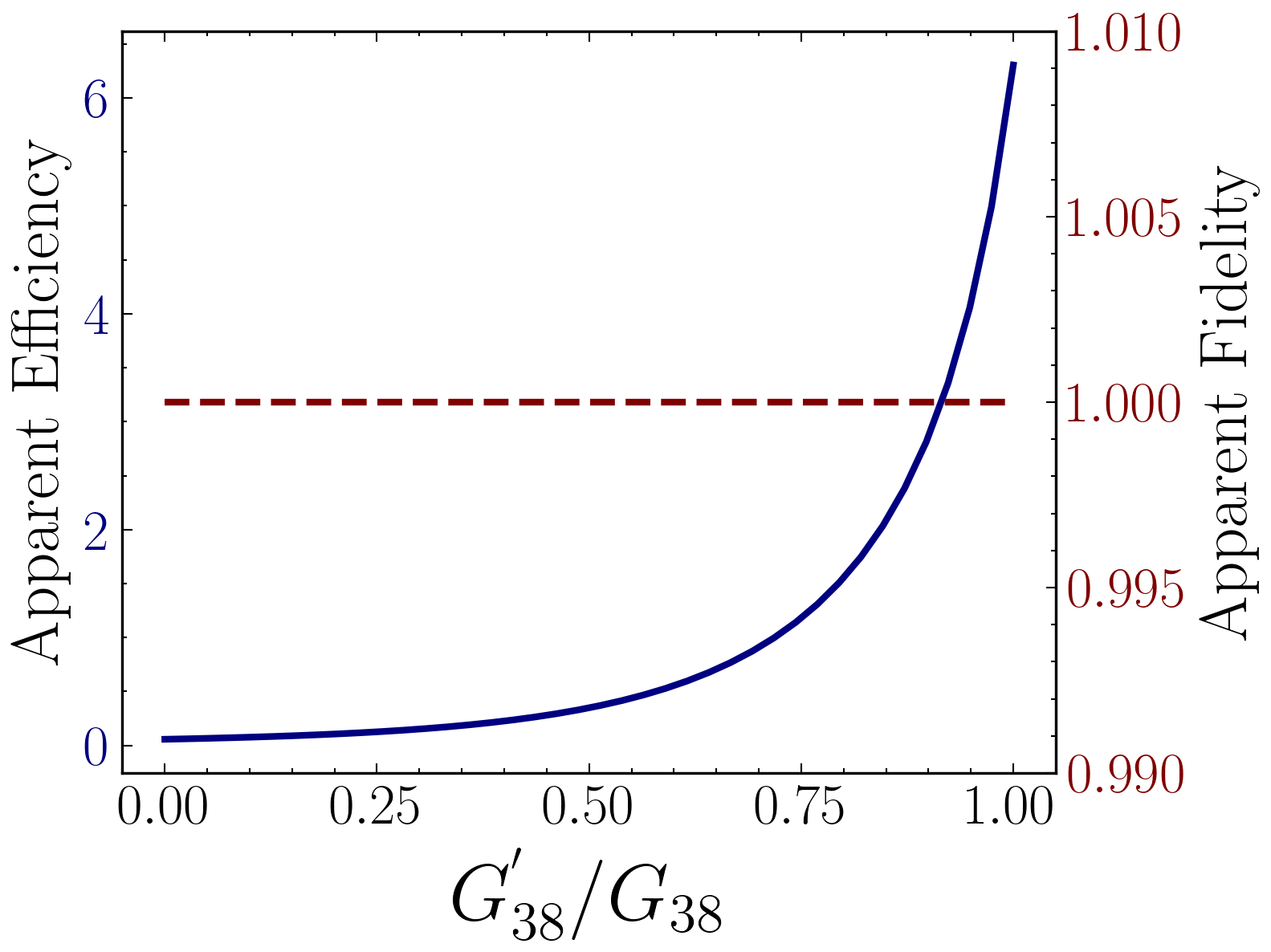} }
{\includegraphics[width=5.8cm]{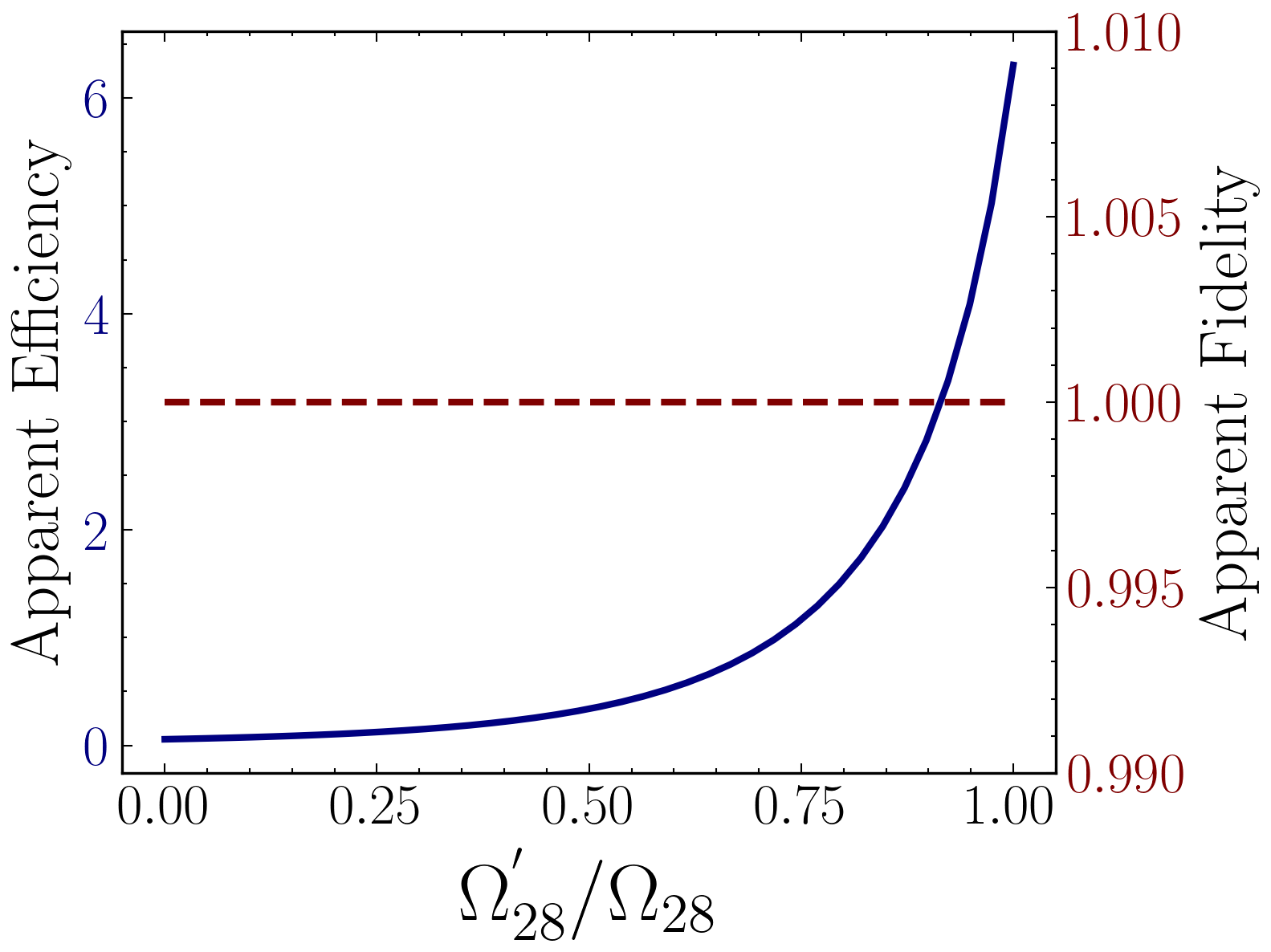} }
{\includegraphics[width=4.9cm]{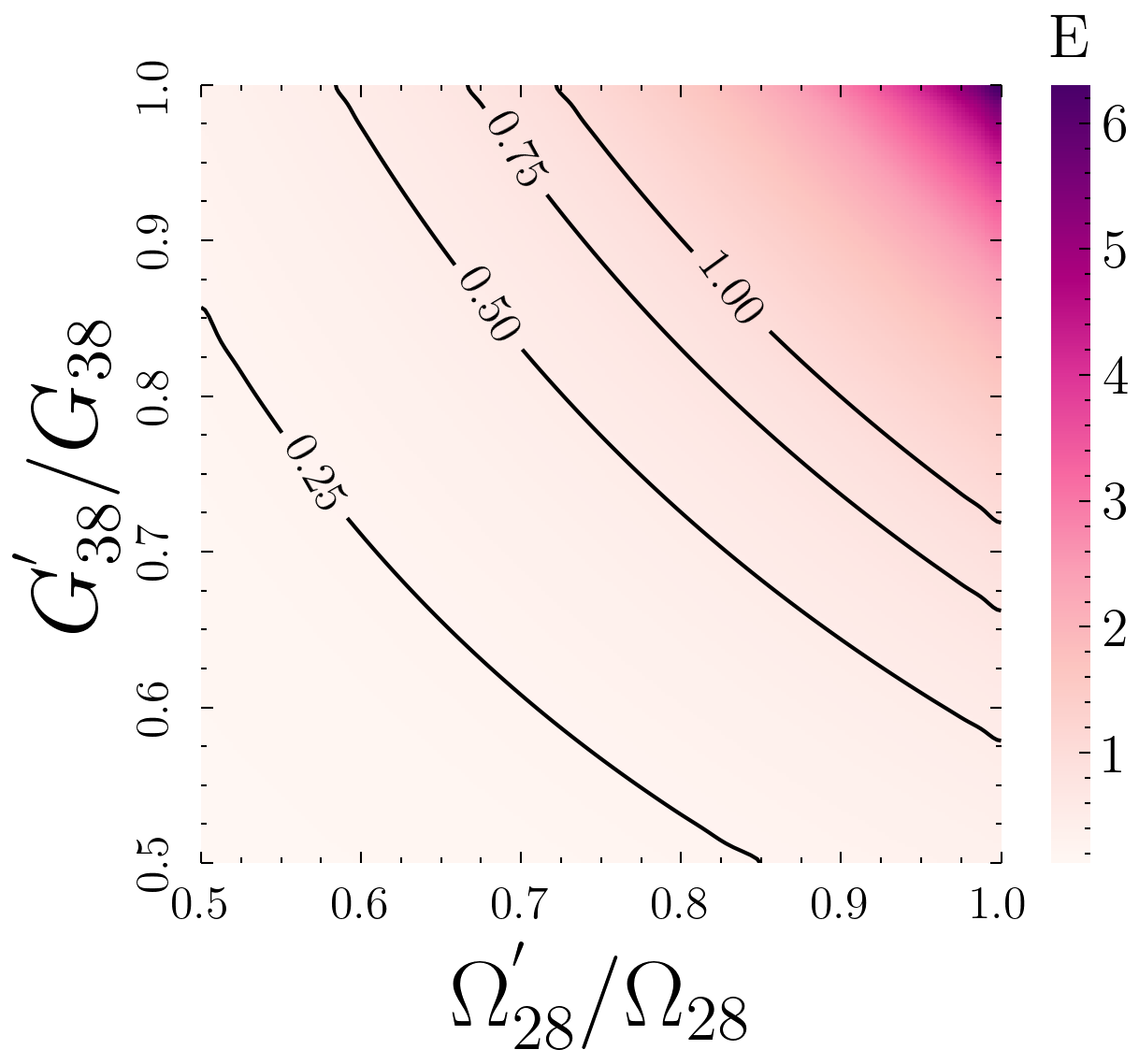} }\\
    \caption{Semi-analytical results: Apparent efficiency (solid line) and fidelity (dashed line) of the 4-level system as a function of $G_{38}$ and $\Omega_{28}$, with all other unwanted couplings set to zero. All parameters are consistent with those used in Figure 4 of the paper.}%
    \label{semir}
\end{figure*}
\subsection{Efficiency oscillation with storage time}\label{app:oscillation}
In addition to amplification, as shown in \cref{oscillation}, we observe that when unwanted couplings are included, the apparent memory efficiency oscillates with storage time. This behavior arises because certain unwanted terms in the equations of motion are multiplied by $\text{exp}(\pm i \delta t)$, with $t$ representing the total duration of the process, including the time required to apply both the first and second control fields. The second control field is applied only after the storage time, meaning that changes in the storage time affect $t$. 
During the storage time, the population does not oscillate between levels $\ket{2}$ and $\ket{3}$, indicating an absence of dynamic changes during this period.
Instead, non-trivial interference between the desired and undesired terms introduces a phase shift between these levels, which 
influences the retrieval process. Consequently, the apparent efficiency oscillates with a period of $\delta/\pi$. On the other hand, when all unwanted couplings are eliminated, the efficiency consistently remains below unity as shown in \cref{oscillation}.

 In general, the apparent efficiency oscillation amplitude depends on both the magnitude and direction of the external fields, which affect the strength of the desired and undesired couplings as well as the splitting between the energy levels. To make the amplitude of oscillations negligible, one can adjust the direction of the external fields to minimize $\delta$. Consequently, in systems with degenerate ground states (i.e., $\delta = 0$), efficiency does not oscillate with storage time. However, the absence of oscillation does not necessarily imply the absence of interference in the system. It should also be noted that the amplification is not due to oscillations in apparent efficiency. In fact, there are operating regimes where oscillations are negligible, yet efficiencies can still exceed unity. For instance, manually setting $\delta = 0$ in the 9-level NV-center system eliminates oscillations, while the efficiency remains constant at a value of $2.09$ for all storage times, as shown in \cref{oscillation}. Therefore, if present, the oscillation can cause additional reduction or enhancement of efficiency at a given storage time.

\subsection{Semi-analytical estimations}

Using the numerical solution of Equation 2 of the paper, the semi-analytical estimation also indicates that the apparent efficiency can exceed unity, while the apparent fidelity remains constant at unity, as shown in \cref{semir}.
 Here, the apparent efficiency exceeds that obtained from the numerical estimation of the 9-level system. This difference can be attributed to (i) the presence of additional unwanted couplings in the 9-level case (see Figure 1 of the paper), which can interfere destructively, (ii) adjustments in the amplitudes of the control fields, and (iii) the adiabatic elimination assumption used to derive Equation 2. 
The heatmap plot in \cref{semir} shows that reducing the unwanted coupling $G_{38}$ has the same effect on apparent efficiency as reducing the other unwanted coupling, $\Omega_{28}$. 

The last term in Equation 2 is the most significant unwanted contribution to amplification in the system. \cref{term8} shows that when this term is absent, the efficiency remains below unity.

\begin{figure}
\includegraphics[scale=0.48]{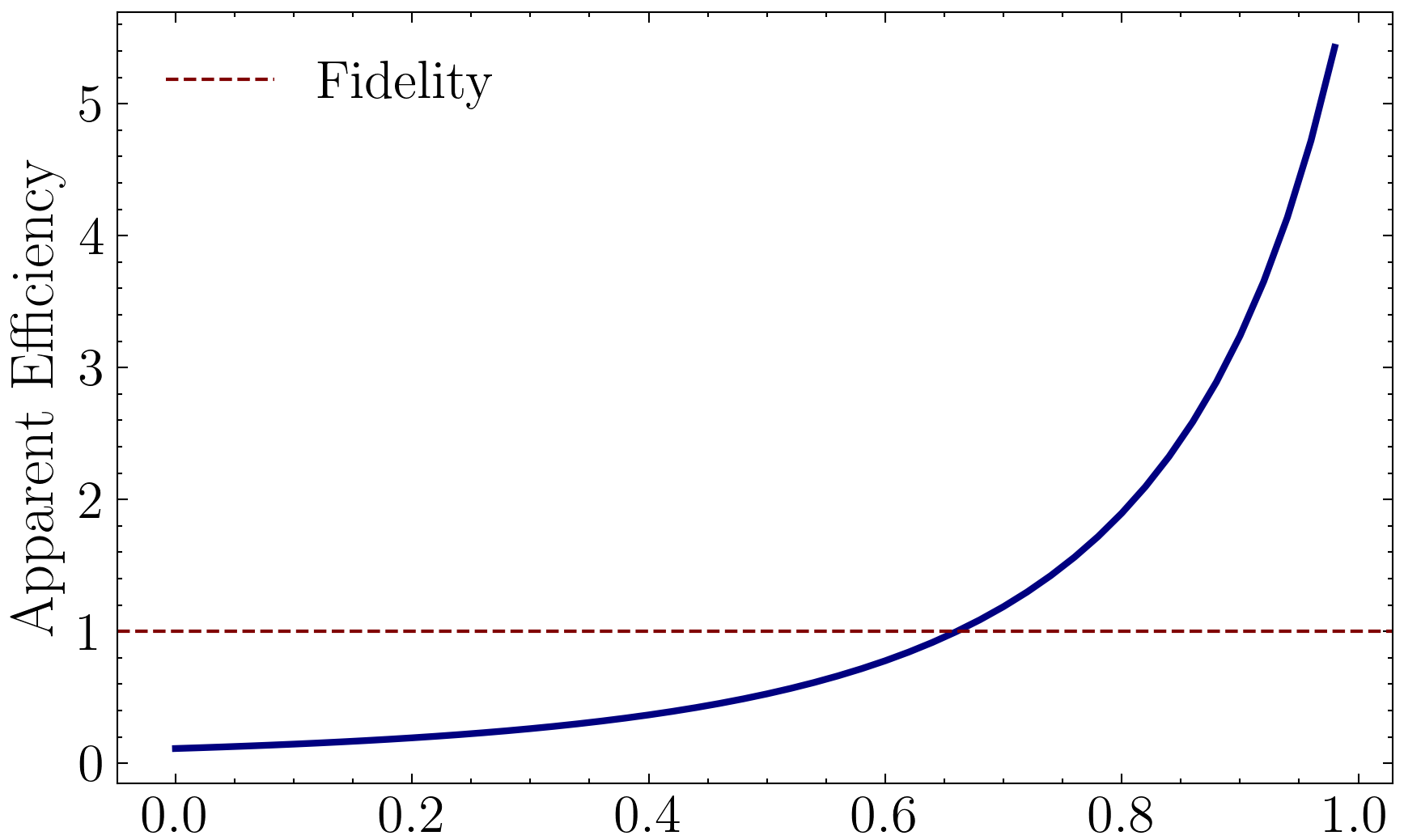}\\

    \caption{Apparent efficiency (solid line) and apparent fidelity (dashed line) of the 4-level system as a function of term 8 in Equation 2 of the paper. Here, all terms except term 8 are fixed at their original values, while term 8 increases from zero to its original value along the x-axis.
    All parameters are consistent with those used in Figure 4 of the paper. }
    \label{term8}
\end{figure}

\begin{table*}
	\caption{List of all possible couplings through the signal field (top) and the control field (bottom). For the parameters used See caption of Figure 2. The desired couplings are highlighted in bold.  }
	\begin{ruledtabular}
		\begin{tabular}{c c c c c c c}
			 &$k=4$&$k=5$&$k=6$& $k=7$&$k=8$&$k=9$\\[0.05cm]
			\hline \\[-0.2cm]
			
			$j=1$&$2.51i$ (Hz)&$- 14.93i$ (Hz)&$2.23i$ (MHz)&$3.66$ (GHz)&$4.21$ (KHz)&$-0.214$ (GHz) \\[0.05cm]

			$j=2$ &$-26.78i$ (KHz)&$-97.19i$ (KHz)&$92.86i$ (MHz)&$0.214$ (GHz)&$5.35$ (MHz)&$\bold{3.66}$ (GHz) \\[0.05cm]
			
			$j=3$ &$-18.34i$ (MHz)&$-66.75i$ (MHz)&$-0.135i$ (MHz)&$-0.316$ (MHz)&$3.67$ (GHz)&$-5.34$ (MHz) \\[0.05cm]
			\hline\hline \\[-0.2cm]
   
$j=1$&$6.77i$ (GHz)&$-1.64i$ (GHz)&$3.34$ (KHz)&$-3.19$ (Hz)&$4.02$  (MHz)&$24.5$ (Hz) \\[0.05cm]

$j=2$&$- 1.64i$ (GHz)&$-6.77i$ (GHz)&$10.3i$ (MHz)&$-21.2$ (KHz)&$-0.131$ (GHz)&$-0.258$ (MHz) \\[0.05cm]	
			
$j=3$&$2.41i$ (MHz)&$9.97i$ (GHz)&$6.97i$ (GHz)&$-14.5$ (MHz)&$0.194$ (MHz)&$\bold{-0.176}$ (GHz) \\[0.05cm]
		\end{tabular}
	\end{ruledtabular}
	\label{tab:couplings}
\end{table*}

\subsection{Additional plots}\label{app:Ad-plots}

\subsubsection{NV center}

\cref{4ln} illustrates the achievable apparent memory efficiency and fidelity in a simplified 4-level system, solved numerically, where only two unwanted couplings, $G_{38}$ and $\Omega_{28}$, are present. Even in this reduced setup, the system exhibits significant amplification, exceeding that seen in the more complex 9-level system. This difference can be attributed to the presence of additional unwanted couplings in the 9-level case (listed in \cref{tab:couplings}), which can destructively interfere with each other. Moreover, as shown in \cref{semih}. b for the numerical 4-level case, a symmetry emerges (similar to the semi-analytical case) in the impact of $G_{38}$ and $\Omega_{28}$ on apparent efficiency; that is reducing $G_{38}$ has the same effect as reducing $\Omega_{28}$, and the apparent fidelity remains exactly unity. 

In all cases (namely, the numerical 9-level system, the numerical simplified 4-level system, and the semi-analytical model) an increase in $\Delta_8$ reduces the apparent efficiency and consequently lowers the amplification by increasing the separation between levels $k=8$ and $k=9$.  Specifically, in the semi-analytical model, increasing $\Delta_8$ has the same effect as reducing $G_{38}$ or $\Omega_{28}$ by the same factor (see \cref{semih}. c). Therefore, to minimize amplification in this model, the ratio $G_{38} \Omega_{28} / \Delta_8$ should be minimized. 
 
It should be noted that although, in this paper, the direction and magnitude of the external fields are chosen to establish the intended optical polarization selection rules, as in Ref. \cite{heshami2014raman}, we have identified several other operating regimes where the apparent memory efficiency exceeds unity while the apparent noise remains zero. As such, while adjusting the direction and magnitude of external fields to minimize the ratio of undesired to desired couplings may enhance memory performance,  finding a regime where (i) all undesired couplings are negligible compared to the desired ones and (ii) both efficiency and fidelity are sufficiently high for quantum applications may not be feasible within the energy level structure considered in this study. Therefore, exploring an alternative set of levels to define the primary $\Lambda$ system may be advantageous.

\begin{figure*}
    \centering
{\includegraphics[width=5.7cm]{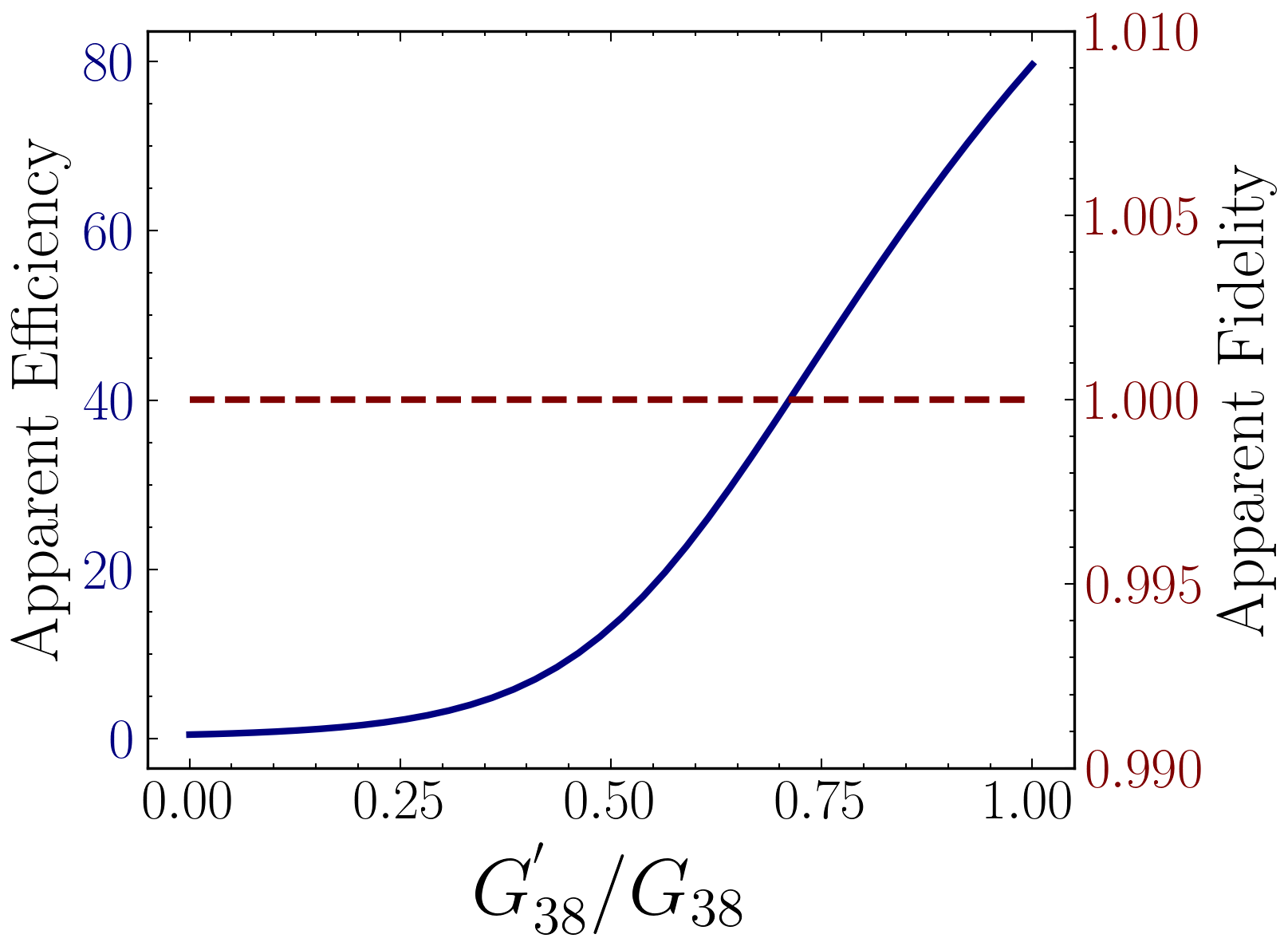} }
{\includegraphics[width=5.7cm]{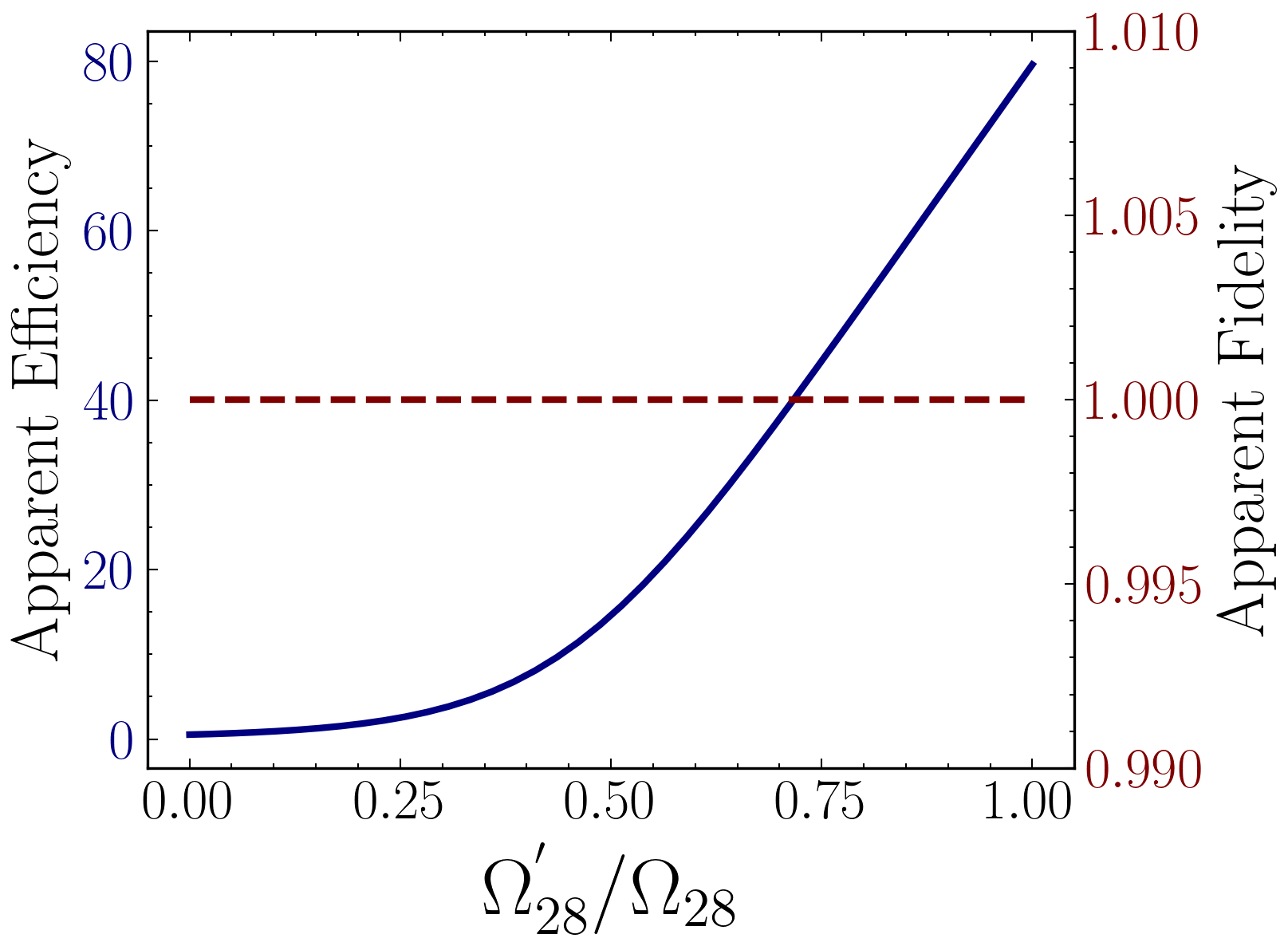} }
{\includegraphics[width=4.8cm]{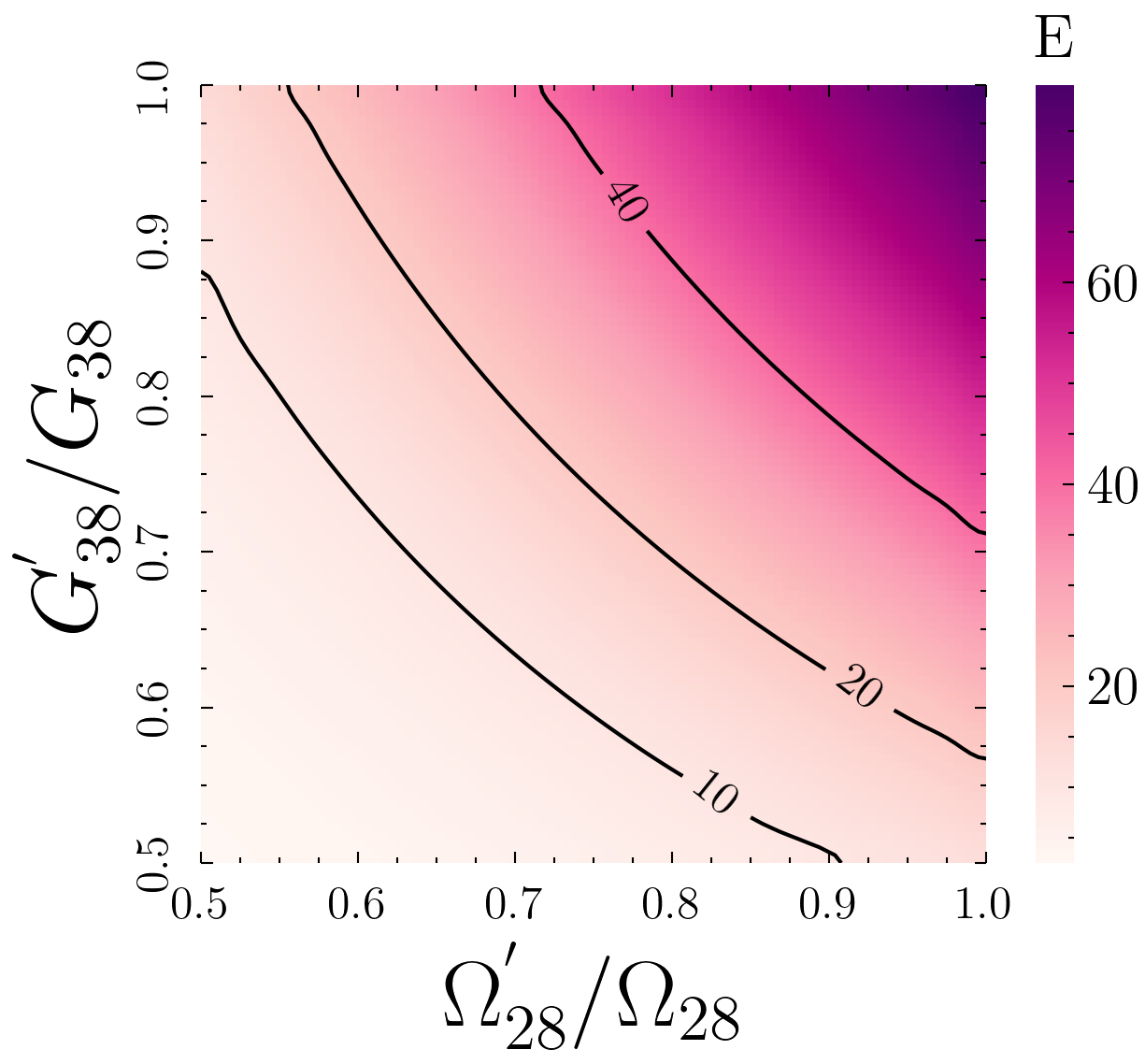} }
    
    \caption{Numerical results: The apparent efficiency (solid line) and fidelity (dashed line) of the simplified 4-level system are presented as a function of $G_{38}$, and $\Omega_{28}$. The heatmap illustrates how apparent efficiency varies as a combined function of $G'_{38}$ and $\Omega'_{28}$.}%
\label{4ln}
\end{figure*}

\begin{figure*}
    \centering

\hspace{-150mm}\textbf{(a)}\\
\hspace{-4mm}\includegraphics[scale=0.45]{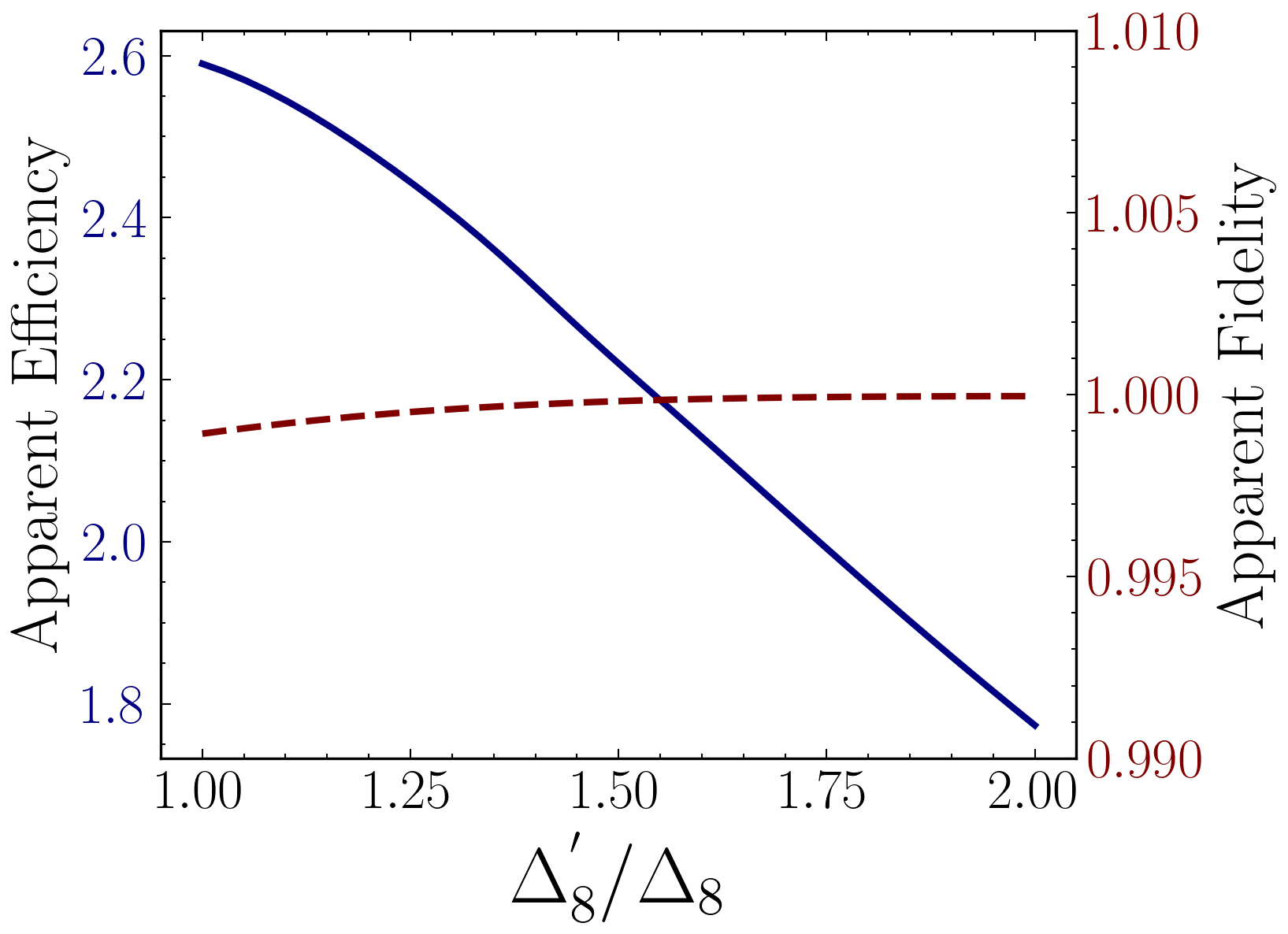}\includegraphics[scale=0.5]{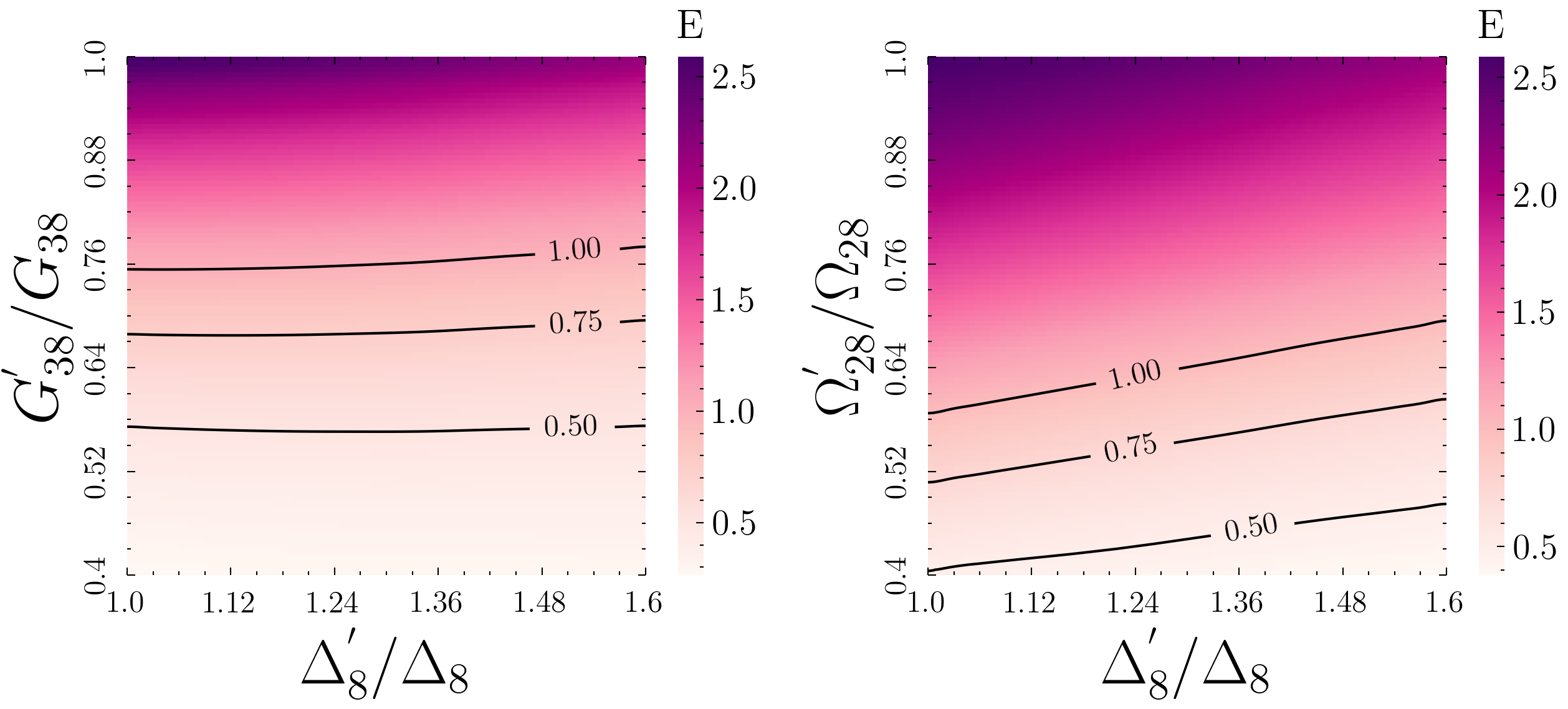}\\
     \hspace{-150mm}\textbf{(b)}\\
     \hspace{-4mm}\includegraphics[scale=0.45]{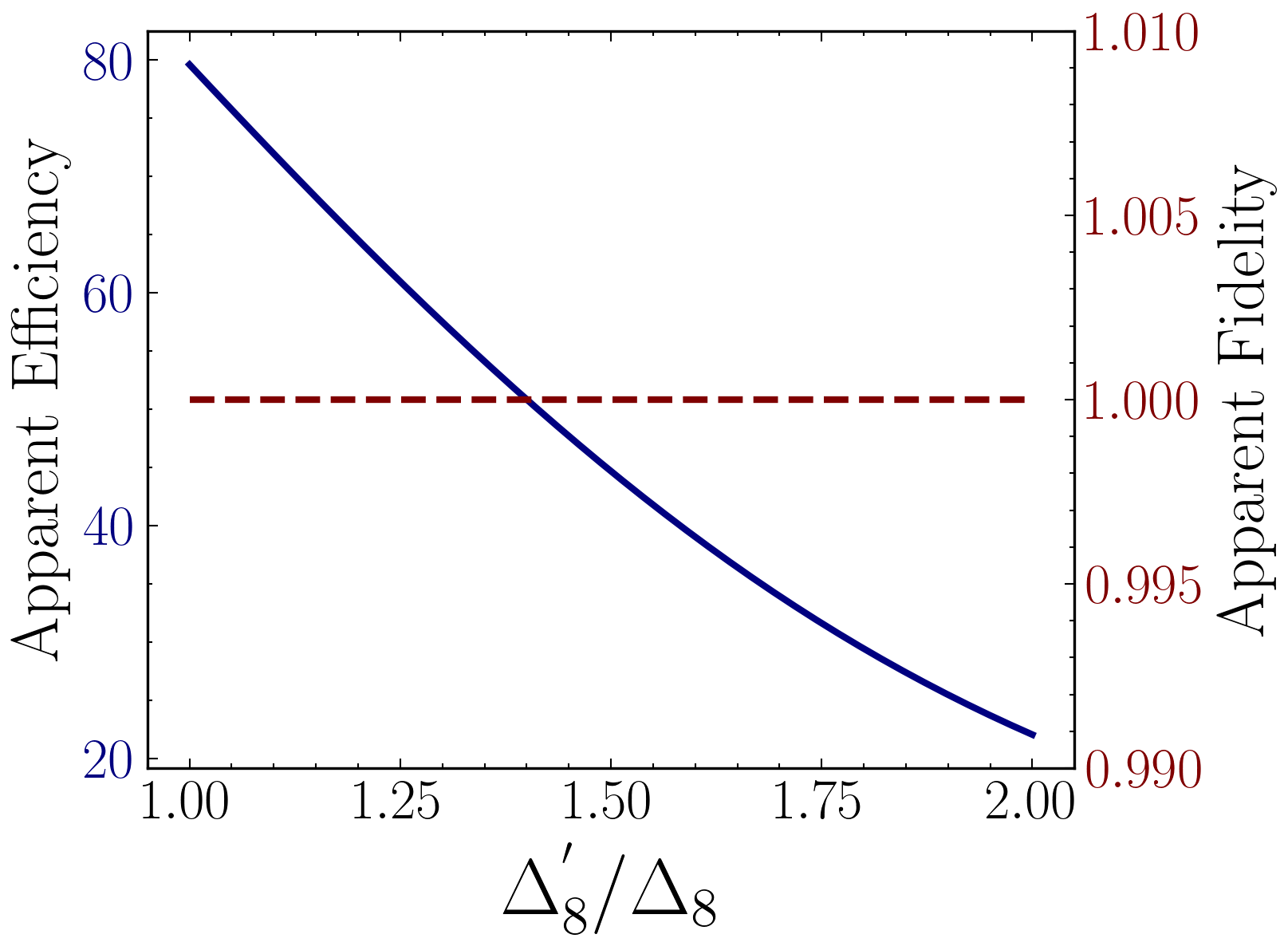}
     \includegraphics[scale=0.5]{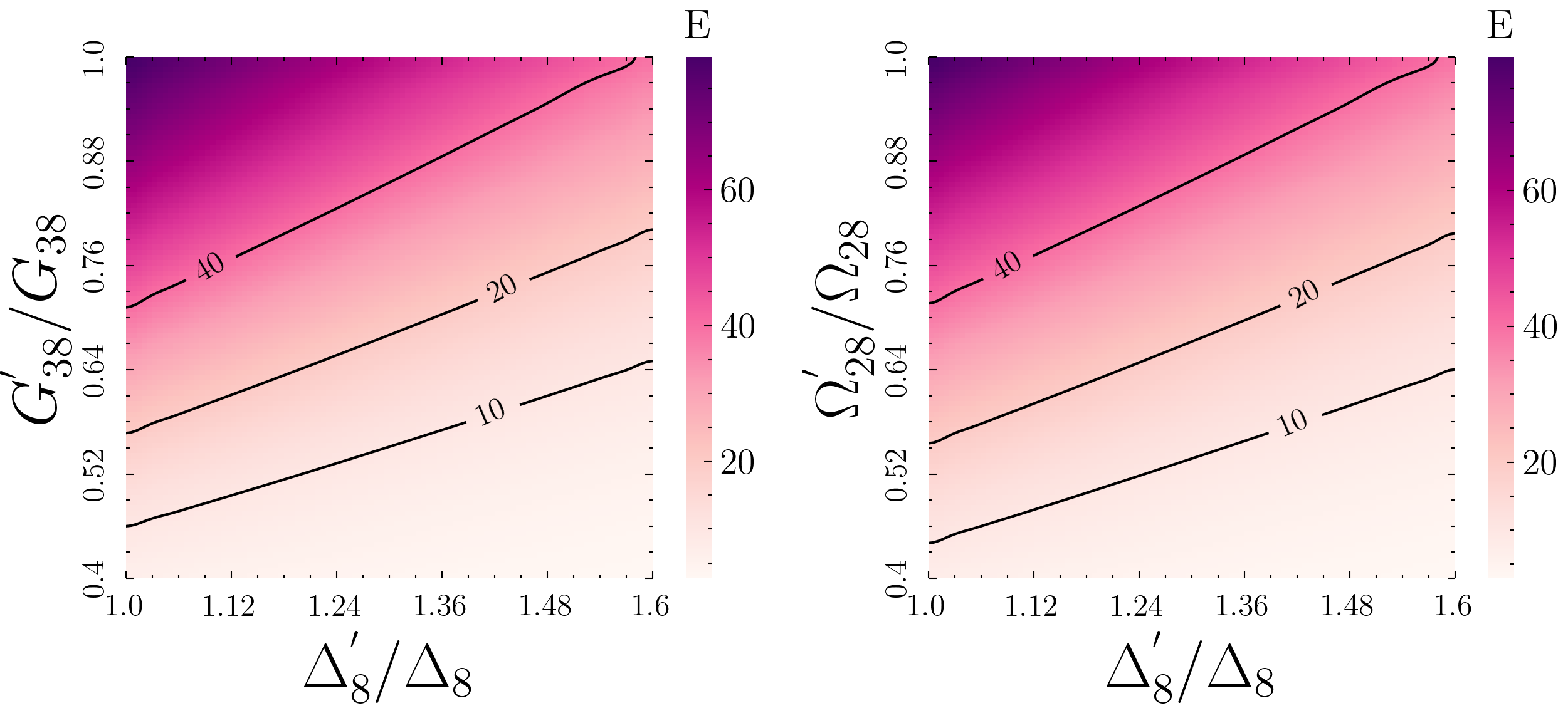} \\
     \hspace{-150mm}\textbf{(c)}\\
    \hspace{-5.mm}\includegraphics[scale=0.45]{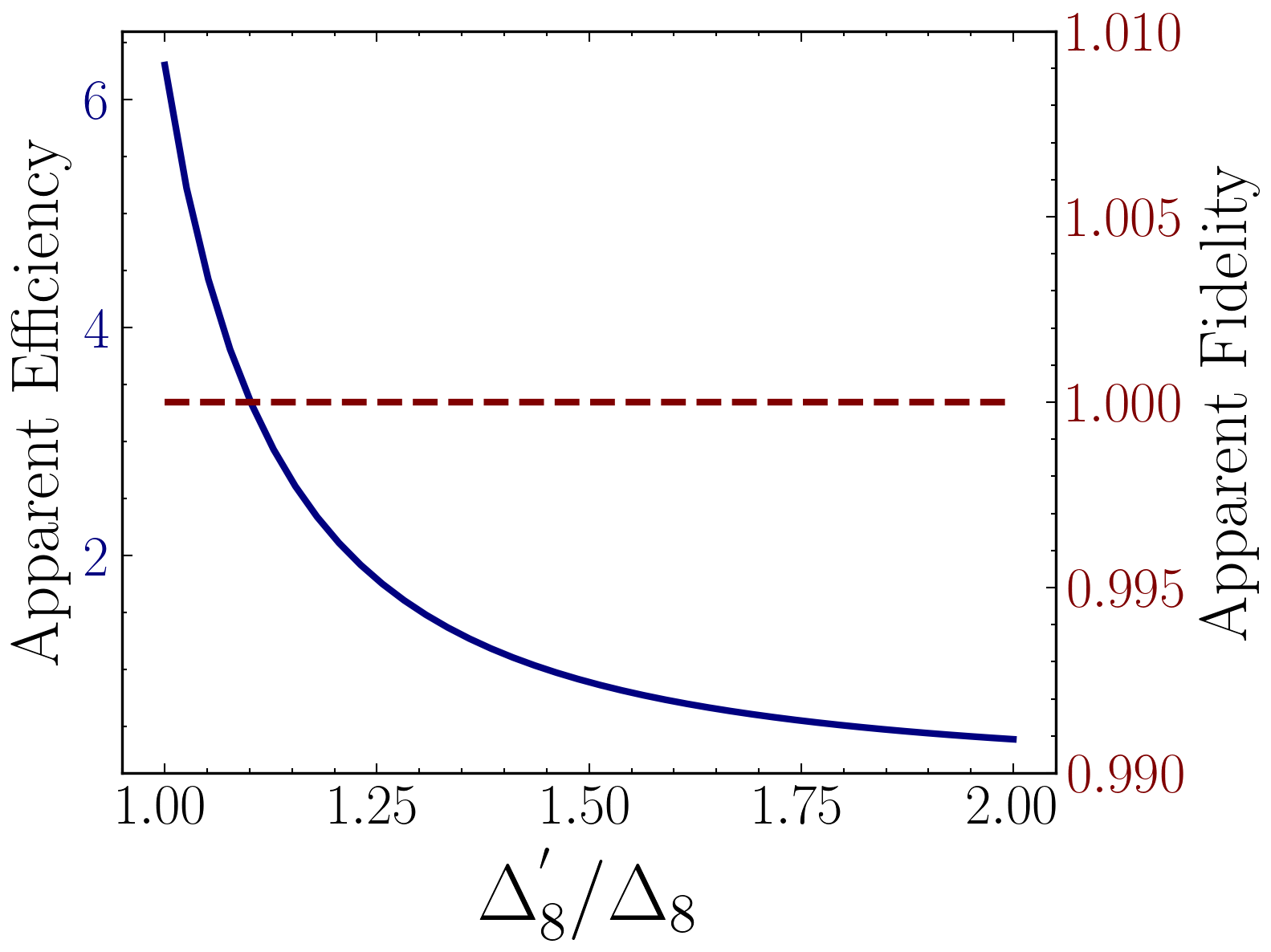}\includegraphics[scale=0.5]{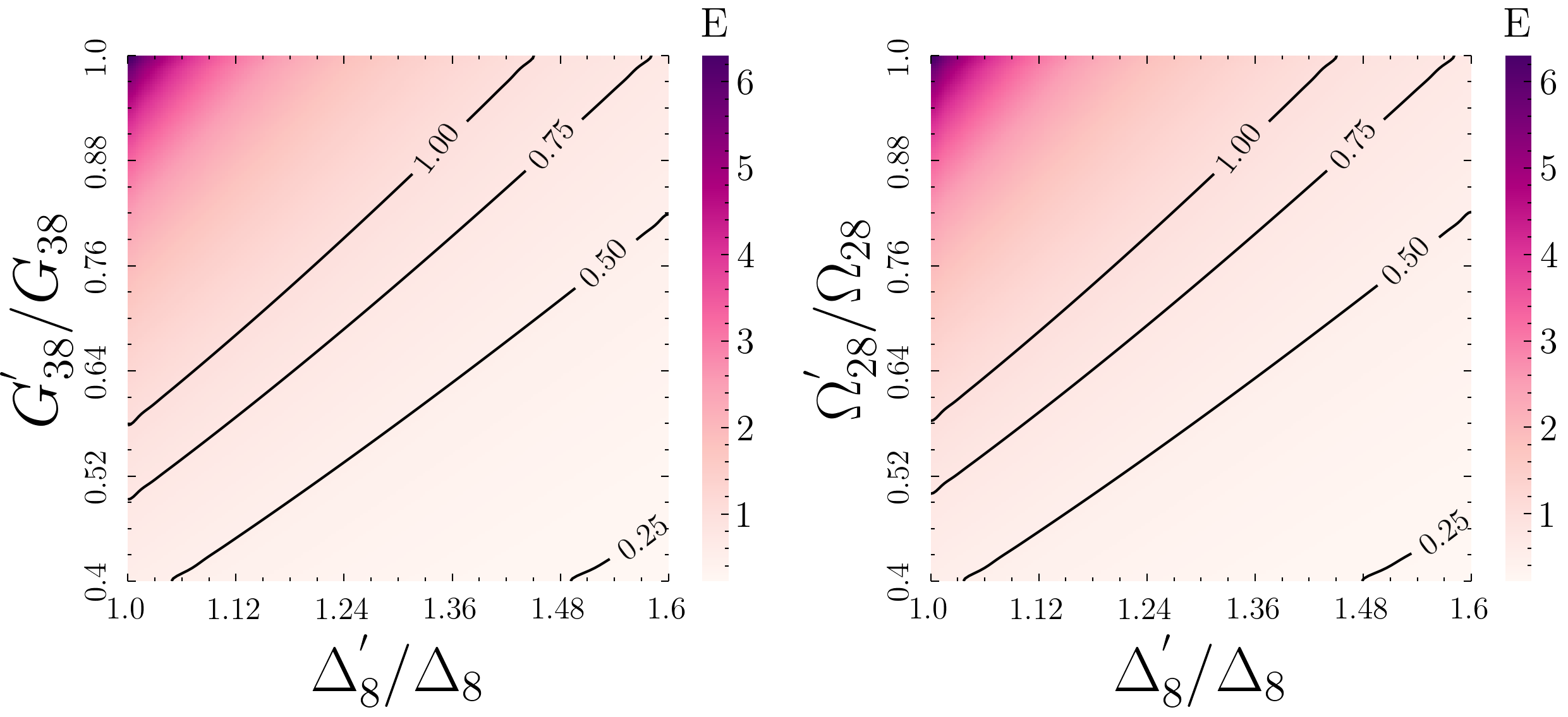}\\\vspace{-01.0mm}

\caption{Variation in apparent fidelity (dashed-line) and efficiency (solid-line) with respect to the splitting $\Delta_8$ for the \textbf{(a)} 9-level system, \textbf{(b)} simplified 4-level system, and \textbf{(c)} semi-analytical case. The heatmaps illustrate how apparent efficiency varies as a combined function of $\Delta_8$ (with the original value of 1.6 GHz) and the unwanted couplings.}
    \label{semih}
\end{figure*}

\begingroup
\subsubsection{Rubidium memory}

The energy levels of the $D_1$ line of the $^{87}$Rb are shown in \cref{fig:Rb}, featuring two hyperfine ground states and two hyperfine excited states. Using the transition rules for the three different polarizations i.e., $\pi$ (linear) and $\sigma^\pm$ (circular), one can compute the dipole matrix elements for all these levels. Notably, the values remain the same across these polarizations, as listed in \cref{tab:HVcouplings} \cite{steck2001rubidium}. Since this holds for different polarizations, the polarization of the signal and control beams does not affect the results. Here, the Zeeman levels are not resolved so the atoms are distributed in all magnetic sublevels.

The actual values of the transition dipole elements are obtained by multiplying the values in the table above by $d_z = \braket{J = \frac{1}{2} | d | J' = \frac{1}{2}}$,  
which corresponds to the transition dipole moment of the $D_1$ line, with a value of $2.537 \times 10^{-29} \ \text{C} \cdot \text{m}$ \cite{steck2001rubidium}. Using these dipole moments and the relations  
$G_{jk}=d_z g_x(j,k) \sqrt{\omega_c/2 V \hbar \epsilon}$, and $\Omega_{jk}=d_z g_y(j,k) E_2/2\hbar$, were $\lambda_{ge} = 794.97\,\text{nm}$, and $n_d = 1$,
one can estimate the coupling strength of  $^{87}$Rb gas.

\cref{hvdelta} illustrates how apparent efficiency and fidelity change with variations in the splitting between the excited levels. The heatmaps also show the relationship between unwanted couplings and the $\Delta_8$ splitting.
\begin{table}[h]
\centering
\caption{Transition dipole elements of $D_1$ line for $^{87}$Rb.}
\fbox{  
\begin{tabular}{c | c c}  
    & $F' = 1$ & $F' = 2$ \\[0.05cm]
    \hline \\[-0.2cm]
    $F = 1$\,\,\, & \,\,\,$g(1,1) = \sqrt{1/6}$ &\,\,\,\,\,\, $ g(1,2) = \sqrt{5/6}$ \\[0.075cm]
    $F = 2$\,\,\, &\,\,\, $g(2,1) = \sqrt{5/6}$ & \,\,\,\,\,\,$ g(2,2) = \sqrt{5/6}$  \\[0.05cm]
\end{tabular}
}  
\label{tab:HVcouplings}
\end{table}

\begin{figure}
	\centering
	\includegraphics[width=7cm]{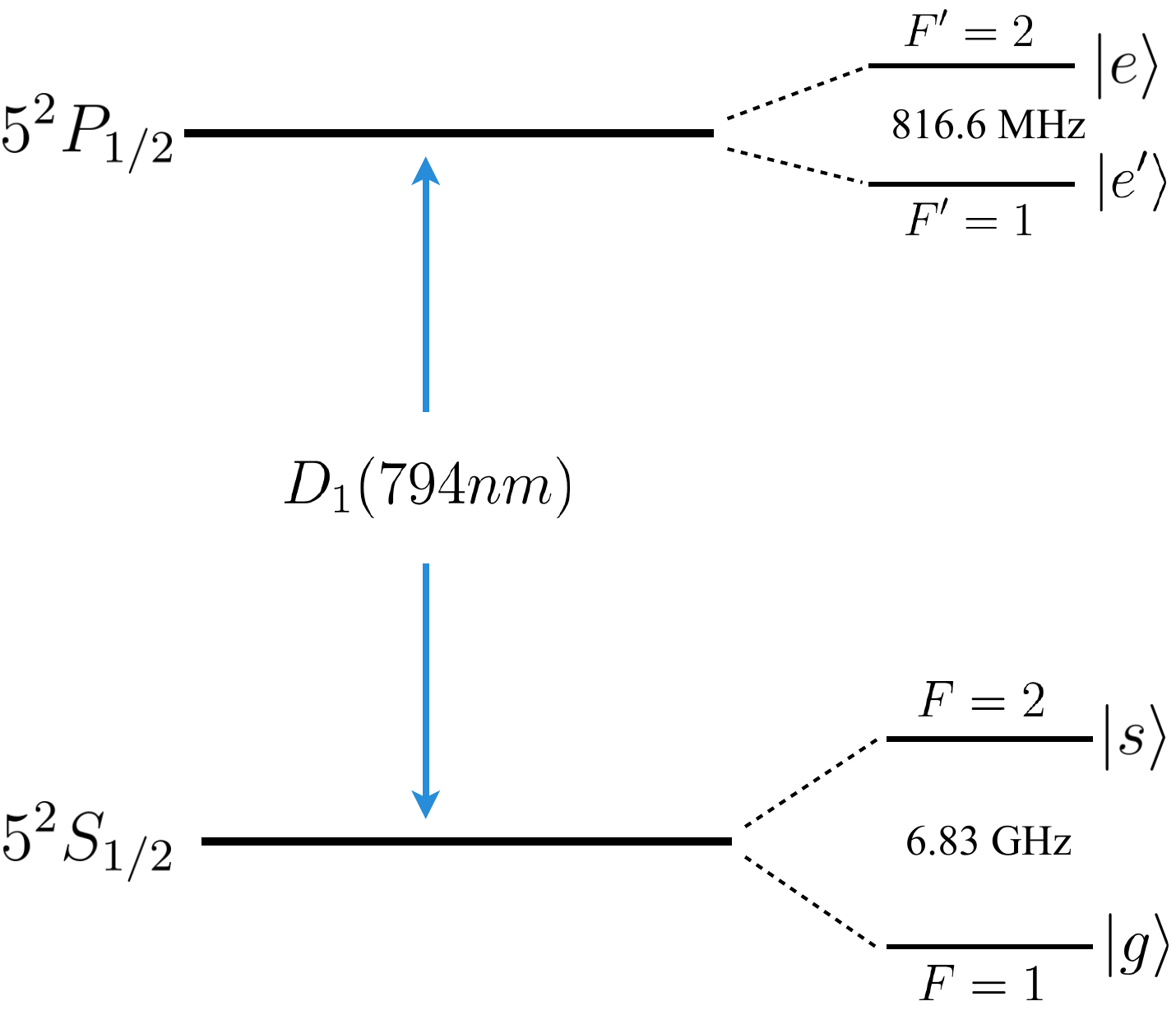}
	\caption{Energy level structure of the $^{87}$Rb in the absence of an external magnetic field. Here, $F$ is the quantum number of the atomic angular momentum.}\label{fig:Rb}
\end{figure}

\begin{figure*}
    \centering
{\includegraphics[width=17cm]{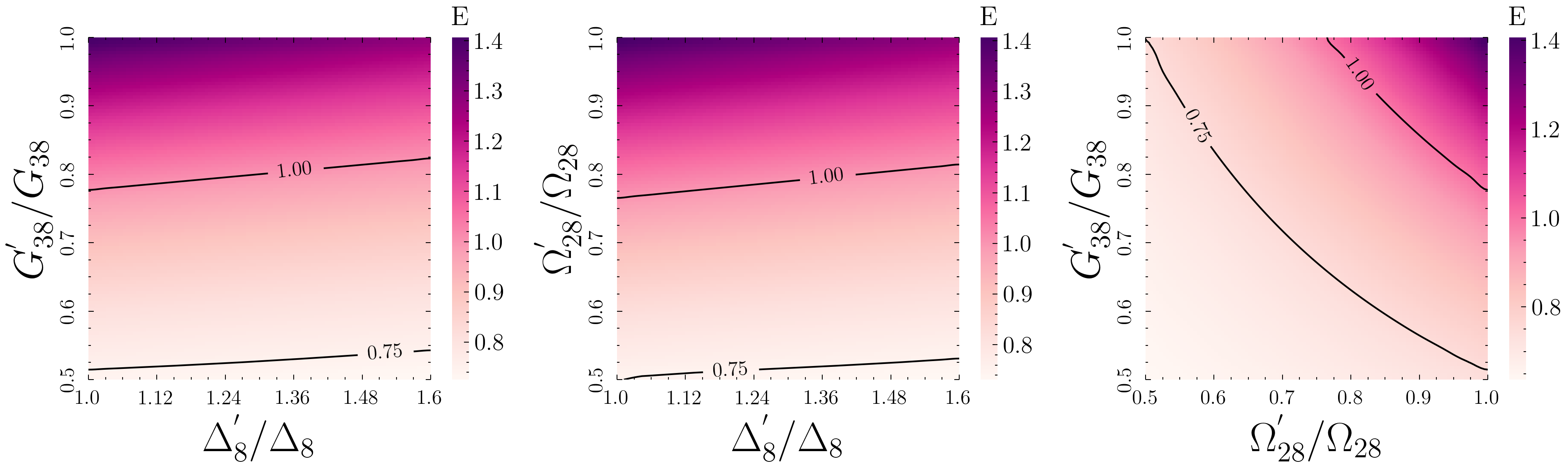} }
\caption{Variation in apparent efficiency as a combined function of $\Delta_8$ and the unwanted couplings for the $^{87}$Rb memory.}
    \label{hvdelta}
\end{figure*}

\end{document}